\documentclass[11pt,a4paper]{article}                    
\usepackage{graphicx}

\usepackage{amsmath}
\usepackage{amssymb}
\usepackage{a4wide}

\newcommand{\BE}{\begin{equation}}
\newcommand{\EE}{\end{equation}}

\newcommand{\cA}{\mathcal{A}}
\newcommand{\cB}{\mathcal{B}}
\newcommand{\cC}{\mathcal{C}}
\newcommand{\cL}{\mathcal{L}}
\newcommand{\cT}{\mathcal{T}}
\newcommand{\cH}{\mathcal{H}}
\newcommand{\cO}{\mathcal{O}}
\newcommand{\cQ}{\mathcal{Q}}
\newcommand{\cK}{\mathcal{K}}

\newcommand{\mR}{\mathbb{R}}
\newcommand{\e}{\mathrm{e}}
\newcommand{\eps}{\epsilon}
\newcommand{\vphi}{\varphi}
\newcommand{\deltaV}{{\delta V}}
\newcommand{\as}{n}
\newcommand{\rL}{{\rm L}}
\newcommand{\JJ}{\boldsymbol{J}}
\newcommand{\PP}{\boldsymbol{P}}
\newcommand{\jj}{\boldsymbol{j}}
\newcommand{\xx}{{\boldsymbol{x}}}
\newcommand{\pp}{{\boldsymbol{p}}}
\newcommand{\vv}{\boldsymbol{v}}
\newcommand{\zz}{\boldsymbol{z}}
\newcommand{\eell}{\boldsymbol{\ell}}
\newcommand{\Set}{\Theta}
\newcommand{\bk}[1]{{\langle #1 \rangle}}
\newcommand{\abs}[1]{{\vert {#1} \vert}}
\newcommand{\norma}[1]{{\Vert {#1} \Vert}}
\newcommand{\pt}{\partial}
\newcommand{\weq}{w^\mathrm{eq}}
\newcommand{\feq}{f^\mathrm{eq}}
\newcommand{\IN}{\mathrm{in}}
\newcommand{\OUT}{\mathrm{out}}
\DeclareMathOperator{\DIV}{div}
\DeclareMathOperator{\sign}{sign}

\newtheorem{theorem}{Theorem}[section]
\newtheorem{lemma}[theorem]{Lemma}
\newtheorem{proposition}[theorem]{Proposition}

\newtheorem{definition}[theorem]{Definition}
\newtheorem{remark}[theorem]{Remark}

\numberwithin{theorem}{section}

\begin{document}

\title{Quantum transmission conditions for diffusive transport in graphene with steep potentials}

\begin{center}
\Large{\textbf{Quantum transmission conditions for diffusive transport in graphene with steep potentials}}
	\\[.5cm]
	\small{L. Barletti}	\\
   \vskip0.2cm
	\textit{\textsf{Dipartimento di Matematica e Informatica ``U. Dini'' \\
Viale Morgagni 67/A, I-50134 Firenze, Italia}}	
\vskip0.2cm
	\textit{luigi.barletti@unifi.it}
	\\[.5cm]	
		\small{C. Negulescu}	\\
\vskip0.2cm
	\textit{\textsf{Institut de Math\'ematiques de Toulouse, Universit\'e  Paul Sabatier\\
118, Route de Narbonne
F-31062 Toulouse, France}}	
\vskip0.2cm
	\textit{claudia.negulescu@math.univ-toulouse.fr}
\end{center}

\author{Luigi Barletti         \and
        Claudia Negulescu
}

\begin{abstract}
We present a formal derivation of a drift-diffusion model for stationary electron transport in graphene, 
in presence of sharp potential profiles, such as barriers and steps.
Assuming the electric potential to have steep variations within a strip of vanishing width on a macroscopic scale, 
such strip is viewed as a quantum interface that couples the classical regions at its left and right sides.
In the two classical regions, where the potential is assumed to be smooth, electron and hole transport is described in terms of semiclassical kinetic equations.
The diffusive limit of the kinetic model is derived by means of a Hilbert expansion and a boundary layer analysis, and consists of drift-diffusion 
equations in the classical regions, coupled by quantum diffusive transmission conditions through the interface.
The boundary layer analysis leads to the discussion of a four-fold Milne (half-space, half-range)  transport problem.

\end{abstract}

\section{Introduction}
\label{S1}
%
Theoretical prediction and experimental demonstration of striking quantum phenomena manifested by electrons in graphene, such as Klein paradox
\cite{KatsnelsonEtAl06,Young09} and Veselago lensing \cite{CheianovEtAl2007,Lee15}, are among the most important achievements of solid-state physics in
the last decade, and offer interesting opportunities to nano-electronics and opto-electronics. 
All such phenomena are intimately related to the chiral nature of electrons in graphene \cite{CastroNeto09} and take place in presence of electric
potential steps or barriers, that can be realised by means of suitable electric gates or doping profiles.
On the other hand, such effects depend on the quantum coherence of the electrons and their neat manifestation is only possible in idealised situations, 
or at least in very controlled experimental settings, where the transport is essentially ballistic.
Collisional and diffusive transport, instead, is a more realistic regime in ordinary conditions \cite{Borysenko10,CastroNeto09}
but tends to increase the decoherence, which results in blurred versions of the purely ballistic pictures.
It is therefore important to offer a mathematical instrument to describe and analyze such more realistic situation.
\par
We propose here a hybrid model where a thin ``active'' quantum region, containing to the rapid potential variations, is viewed as a ``quantum interface''
that couples the surrounding ``classical'' regions, where the transport regime is diffusive and incoherent.
The coupling is firstly described at the kinetic level, where the classical-quantum matching is more natural, and then the diffusive limit is performed
by means of the Hilbert expansion method \cite{Degond_review}.
This first, theoretical paper is devoted to the derivation of the model, which will be numerically tested in a subsequent work.
We remark that part of the contents of the present paper have been anticipated in Ref.\ \cite{BarlettiNegulescu17}.
\par
A hybrid kinetic-quantum model for standard particles (i.e.\ scalar particles with parabolic energy-band, as opposed to chiral particles with conical energy-band, as electrons in graphene) has been firstly considered by Ben Abdallah \cite{NBA98,NBA02}.
The central idea in Ben Abdallah's construction is that a scattering problem is solved in the quantum region,
that is a thin strip around the steep potential variations, and the resulting scattering states (in\-ci\-dent/re\-flec\-ted/trans\-mit\-ted waves) are identified with 
inflow/outflow particles in/from the classical regions.
This leads to a hybrid model where {\em transmission conditions}, of quantum nature, are imposed to classical kinetic 
equations.
\par
The diffusive limit of Ben Abdallah's model is studied by Degond and El Ayyadi in Ref.\ \cite{DEA02}.
Here, the kinetic model of Ref.\ \cite{NBA98} is expanded in powers of the scaled collision time (Hilbert expansion),
which leads to classical Drift-Diffusion (D-D) equations in the classical regions. 
The Hilbert expansion of the kinetic transmission condition yields purely classical diffusive transmission conditions at leading order. 
However, a boundary-layer analysis shows that there is a first-order quantum correction of the diffusive transmission conditions under the form
of an ``extrapolation coefficient'' (somehow analogous to the extrapolation length of neutron transport theory \cite{BSS84}), which depends on the
reflection and transmission coefficients coming from the quantum scattering problem.
\par
An intermediate (between kinetic and diffusive) hybrid classical-quantum model has been studied in Ref.\ \cite{DS98}, where two SHE 
(Spherical Harmonic Expansion) models are coupled via suitable interface conditions.
\par
As explained above, our goal is to construct a diffusive model of the electron transport in a graphene device where a small 
(compared to a macroscopic scale)  region, containing the steep potential variations, is the ``active'' zone where quantum coherence is exploited.
Although our construction is inspired by the quoted works \cite{NBA98,DEA02,DS98}, nevertheless we have to deal here with a rather different situation.
First of all, electrons in graphene have a chirality, which is  an additional, discrete degree of freedom, denoted by $s$; 
this implies that, in each classical region, two populations of electrons (corresponding to $s=1$ and $s=-1$) have to be considered.
The two populations, in absence of other coupling mechanisms in the bulk, are coupled by the quantum interface.
The second aspect is that electrons have a conical dispersion relation (energy band), which requires the use of a semiclassical%
\footnote{According to the terminology adopted, e.g., in \cite{Ashcroft}, we call ``semiclassical'' 
classical transport (or Boltzmann) equation where elements of quantum nature are retained, e.g.\ a non-parabolic 
dispersion relation.}
transport equation and a non-standard Fermi-Dirac (F-D) distribution.
Finally, electrons with negative chirality have a negative energy cone which is unbounded from below; this fact forces us to describe such electrons 
in terms of  holes (electron vacancies). 
This is not a novelty, of course, but the fact that positive-energy and negative-energy electrons are coupled by the quantum interface makes 
the introduction of holes a delicate issue.
\par
The content of the present  paper is the following.
After a brief review of the basic facts about the quantum dynamics of electrons in graphene (Section  \ref{S2}),
the construction of the model begins, at the kinetic level, in Sec.\  \ref{S3}.
If $(x,y)$ are the coordinates on the graphene sheet, we assume that the electric potential is a sum $V(x)+U(x,y)$, 
where $V(x)$ has steep variations within a tiny strip around $x = 0$  and tends to a constant potential difference $\deltaV$ 
outside.
This is the potential which is responsible for the quantum effects and is treated by means of the stationary Schr\"odinger equation.
The second term, $U$ is the smooth part of the potential: it is treated semiclassically and produces the drift term of the D-D equations.
Assuming that the width of the quantum strip vanishes on a macroscopic scale, this picture corresponds to a configuration where $x=0$ is a quantum
interface separating the classical regions $x<0$ and $x>0$ (see Figure \ref{device}). 
In Sec.\ \ref{S3.1} we write down the stationary transport equations in the classical regions and the kinetic transmission conditions (KTC) at $x=0$. 
The KTC express the fact that inflowing/outflowing classical particles correspond to the incoming/outgoing plane waves described by
by the scattering problem across the interface.
In Sec.\ \ref{S3.2}, in view of the diffusive limit, we add to the transport equation a relaxation term towards two local Fermi-Dirac distributions 
(one for each value of chirality).  
In Sec.\ \ref{S3.3}, the KTC are reformulated in terms of electrons and holes  and are proven to conserve the total charge flux across the interface.
\par
Section \ref{S4} is devoted to the diffusive limit of the kinetic model and contains the main results of the paper.
In Sec.\ \ref{S4.1} we study the diffusion limit in the bulk, that is in the classical regions. 
By means of a Hilbert expansion \cite{Degond_review,Poupaud91} in powers of the typical collision time $\tau$,
we obtain semiclassical, stationary, D-D equations for electrons and holes.
In Sec.\ \ref{S4.2} the we expand the KTC.
At leading-order we immediately obtain diffusive transmission conditions (DTC) as a relation between the chemical potential at the two
sides of the interface.
Such leading-order DTC couple electrons and holes but are not ``quantum'',  to the extent that they are independent on the solution 
of the scattering problem.
In Sec.\ \ref{S4.3} it is shown that the introduction of a boundary-layer corrector in the Hilbert expansion is necessary to obtain the first-order DTC.
Such corrector is associated to a system of four Milne (half-space, half-range) transport equation coupled by non-homogeneous KTC.
The mathematical properties of such four-fold Milne problem are stated in Theorem \ref{T1}, which is the first of the two main 
results of the paper.  
The layer analysis leads to the first-order correction to the DTC (Theorem \ref{T2}), which is the second main result of the paper.
The correction is expressed as a relation between left and right chemical potentials and involves the asymptotic densities associated to the 
Milne problem. 
Such densities, which are a generalization of the extrapolation coefficients of Ref.\ \cite{DEA02}, depend on the scattering coefficients
and, therefore, the first-order DTC contain information coming from the quantum physics of the interface.
In Sec.\ \ref{S4.4} it is examined the special case where the F-D distribution is approximated by the Maxwell-Boltzmann distribution.
Finally, in Sec.\ \ref{S5}, we summarize our results by writing down a diffusive model with DTC for a 
prototypical graphene device.
%
\section{Quantum and semiclassical dynamics of electrons in graphene}
\label{S2}
%
We briefly review here some basic facts about the dynamics of electrons in a graphene sheet. 
For an exhaustive introduction to the subject we address the reader to Ref.\ \cite{CastroNeto09}.
\par
Due to its remarkable mechanical, thermal, optical and electronic properties, graphene has attracted a lot of scientific attention in the last years, 
and is thought to have several possible technological applications, as for example in the design of electronic devices.
It is a two-dimensional crystal of carbon atoms, arranged in a honeycomb lattice.
Since every fundamental cell of the associated Bravais lattice contains two carbon atoms, the honeycomb can be decomposed into 
two inequivalent sublattices.
This property implies the existence of two energy-bands having conical intersections at exactly two points (Dirac points) of the reciprocal fundamental 
cell  \cite{CastroNeto09,SW58}.
Assuming that the so-called inter-valley scattering is negligible, one can consider just a single Dirac point and approximately conical energy bands
(Dirac cones) around that point.
Graphene is therefore a zero-gap semi-conductor with linear, rather than quadratic, dispersion relation.
\par
As graphene is a 2-dimensional crystal, we shall use the 2-dimensional variable $\xx = (x,y)$ to identify the electron position. 
In the vicinity of a Dirac point the dynamics of the electron envelope wave-function is determined by the Dirac-like Hamiltonian
\BE 
\label{DIR_HA}
\cH=c\, \PP \cdot \boldsymbol{\sigma} + V\,\sigma_0\,,
\EE
where $c \approx 10^{6} \, \mathrm{m}/\mathrm{s}$ is the Fermi velocity (often indicated by $v_F$ in literature), $\PP = (P_x,P_y) = -i\hbar\nabla$ is the pseudomomentum operator, $V = V(\xx)$ is the potential energy,%
\footnote{We remark that we are using the potential energy $V$ instead of the electric potential $-V/q$ (where $q>0$ is the elementary charge).} 
and $\sigma_0$ as well as the Pauli matrices $\boldsymbol{\sigma}=(\sigma_x,\sigma_y)$ are given by
$$
\sigma_0 = \begin{pmatrix} 1 & 0 \\ 0 & 1\end{pmatrix},
\qquad
\sigma_x =  \begin{pmatrix} 0 & 1 \\  1 & 0\end{pmatrix}
\qquad
\sigma_y =  \begin{pmatrix} 0 & -i \\  i & 0\end{pmatrix}.
$$
The stationary Schr\"odinger equation associated to the Hamiltonian \eqref{DIR_HA} is the following eigenvalue problem: 
\BE 
\label{SSE}
\begin{pmatrix}
V & c\,\left(P_x - iP_y \right)
\\
c\,\left( P_x + iP_y\right) & V
\end{pmatrix}
\begin{pmatrix}
\psi_1 \\ \psi_2
\end{pmatrix}
= E
\begin{pmatrix}
\psi_1 \\ \psi_2
\end{pmatrix},
\EE
where $E$ is the energy eigenvalue.
Since the wave-function $\Psi=(\psi_1,\psi_2)^t$ is a two-component (bi-spinor) wave-function, we can associate to electrons
(besides the usual $1/2$-spin which is neglected here) an additional discrete degree of freedom.
This is the {\em chirality}, which is analogous to photon helicity and is represented by the operator
$$
 S = 
  \frac{1}{\abs{\PP}} 
  \begin{pmatrix}
0 & P_x - iP_y
\\
 P_x + iP_y & 0
\end{pmatrix},
$$
possessing the two eigenvalues $s = 1$ and $s = -1$.
This quantity can be interpreted as the projection of the pseudospin $\boldsymbol{\sigma}$ on the direction of the pseudomomentum.
For constant $V$, it is readily seen that the solution to the stationary Schr\"odinger equation \eqref{SSE} exists for any given $E \in \mR$
and is given by plane-wave-like functions parametrised by $\pp = (p_x,p_y) \in \mR^2$ and $s \in \{+1,-1\}$, namely:
\BE 
\label{planew}
\Psi_{\pp,s}(\xx) =
\begin{pmatrix}  1 \\  s\,e^{i\, \phi} \end{pmatrix} \e^{\frac{i}{\hbar}\pp\cdot\xx},
\EE
where
$$
  \abs{\pp} = \frac{E-V}{c}, \qquad s = \sign(E-V), \qquad \phi = \arg(p_x+i p_y).
$$
Note that:
\begin{enumerate}
\item
 $\Psi_{\pp,s}(\xx)$ is a simultaneous (generalised) eigenvector of $\cH$, $\PP$ and $S$ and, therefore, it corresponds to a state with defined energy, 
 $E$, pseudomomentum $\pp$ and chirality $s$;
\item
 the energy $E$ has a degeneracy corresponding to rotations in the two-dimensional $\pp$-space;
\item
 the sign of $E-V$ is equal to the chirality $s$, which can be interpreted as the pseudospin being parallel ($s=1$) or antiparallel  ($s=-1$) 
 to the wave direction $\pp$.
\end{enumerate}
The energy dispersion relation, i.e.\ the energy as a function of $\pp$ and $s$ when $V=0$,  is, therefore 
\BE
\label{}
  E_s(\pp) = s c \abs{\pp},
\EE
which corresponds to the positive and negative Dirac cones.
Using a slightly sloppy terminology, we shall also refer to $E_s(\pp)$ as the ``energy bands'' of the electron.
\par
In the semiclassical limit, the electron wave function collapses into states of defined  pseudomomentum $\pp = (p_x,p_y) \in \mR^2$ and chirality $s = \pm 1$, 
and the dynamics is described by the Hamiltonian system
\BE
\label{HSys}
 \left\{
 \begin{aligned}
 &\dot \xx = \nabla_\pp E_s(\pp),
 \\[4pt]
 &\dot \pp = -\nabla_\xx V (\xx),
 \end{aligned}
 \right.
\EE
where the energy-band derivatives, 
\BE
\label{sclvel}
\nabla_\pp E_s(\pp) = \frac{sc\pp}{\abs{\pp}},
\EE
are the associated semiclassical velocities.
From a semiclassical point of view, it is apparent that electrons in graphene behave as if they were massless charged particles.
They move with constant speed $c$ and the direction of motion is either parallel to the pseudomomentum, for electrons with positive chirality/energy, 
or anti-parallel, for electrons with negative chirality/energy, 
and the changes of direction are determined by the electric force.
Note that positive and negative electrons are completely decoupled in the semiclassical picture, which corresponds to the absence of quantum interference 
between the two chirality states (see also Ref.\ \cite{09}).
\begin{remark}
In the following, $s = \pm 1$ or $s = \pm$ will be used indifferently. 
\end{remark}
%
\section{Hybrid kinetic-quantum model}
\label{S3}
%
In this section we introduce a hybrid kinetic-quantum model of electron transport on a graphene sheet 
in presence of steep potentials.
By this we mean that the behaviour of the electrons in proximity of an abrupt potential variation
is described by a fully-quantum scattering problem, while, in the regions where the potential is smooth, 
it is described by a semiclassical transport , or ``kinetic'' equation.
Such description will be the starting point of the derivation of a hybrid {\em diffusive}-quantum model, which will be carried out in Sec.\ \ref{S4}.
\par
\begin{figure}[htbp] 
\begin{center}
\includegraphics[width=.8\linewidth]{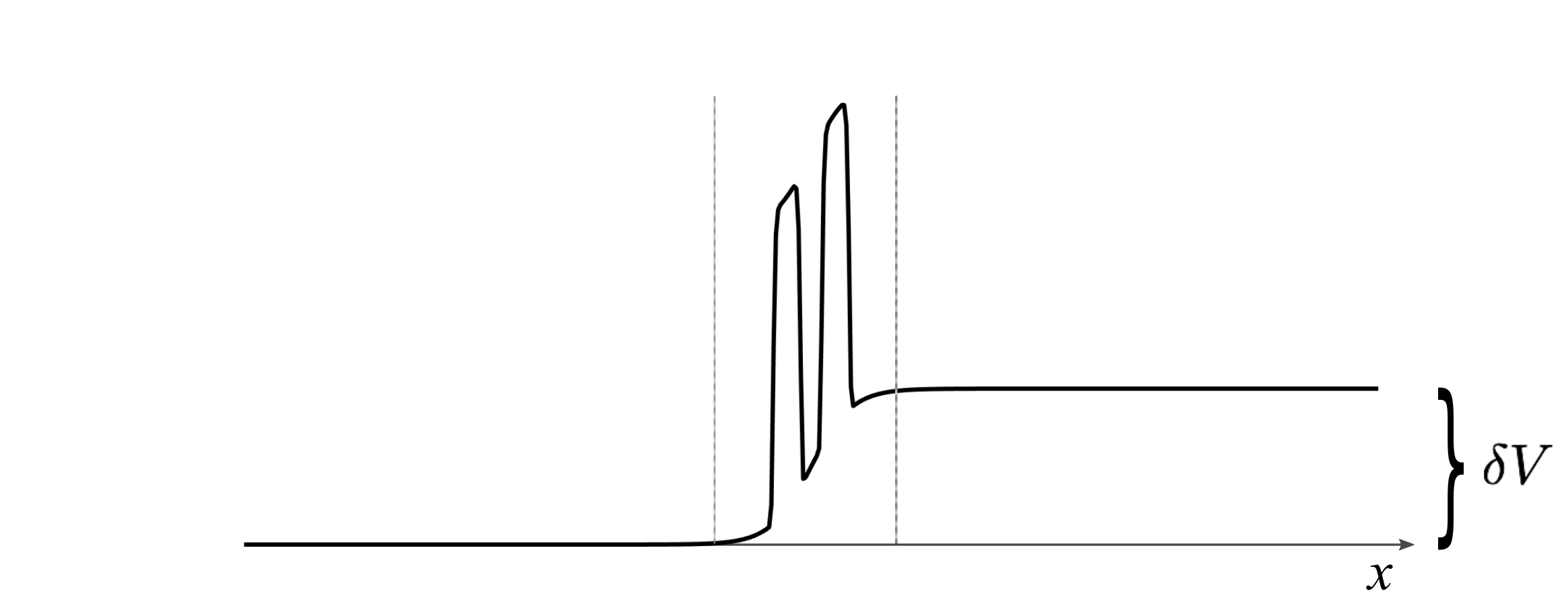}
\end{center}
\caption{Example of an electric potential profile $V(x)$ satisfying assumptions {\bf H1} and {\bf H2} above. 
The ``quantum active region''  lies between the two vertical lines, and $V$ is asymptotically constant with values 0 and $\deltaV$ outside.}
\label{ima_pot} 
\end{figure}

\medskip
For the sake of simplicity, we make the following assumptions on the electric potential energy (see Figure \ref{ima_pot}).
\begin{enumerate}
\item[{\bf H1.}]
$V = V(x)$  depends only on the variable $x$ (which implies that it conserves $p_y$);  
\item[{\bf H2.}]
$V(x) \to 0$ on the left and $V(x) \to \deltaV$  on the right of a ``quantum strip'', around $x = 0$, having
vanishing width on a macroscopic length scale.
\end{enumerate}
For a potential $V(x)$ satisfying {\bf H1} and {\bf H2}, the stationary Schr\"odinger equation \eqref{SSE} has the character of a scattering problem.
In particular, Eq.\ \eqref{SSE} is explicitly solvable outside the quantum-strip and the solutions are recognized to be superpositions of plane waves 
\eqref{planew} with pseudomementum $\pp$ and chirality $s$.
Imposing the continuity of the two-component wave function inside the quantum strip with the outside plane-wave solutions, yields
the scattering  (reflection and transmission) coefficients as functions of energy, which constitute the most relevant information associated 
to the scattering problem.
The fact that the plane-wave exterior solutions have defined $\pp$  and $s$ allows, as explained below, to interpret such 
waves as particles flowing in and out from the classical regions, which permits to match the two classical regions via the reflection and 
transmission coefficients.
\par
\begin{figure}[htbp] 
\begin{center}
\includegraphics[width=.55\linewidth]{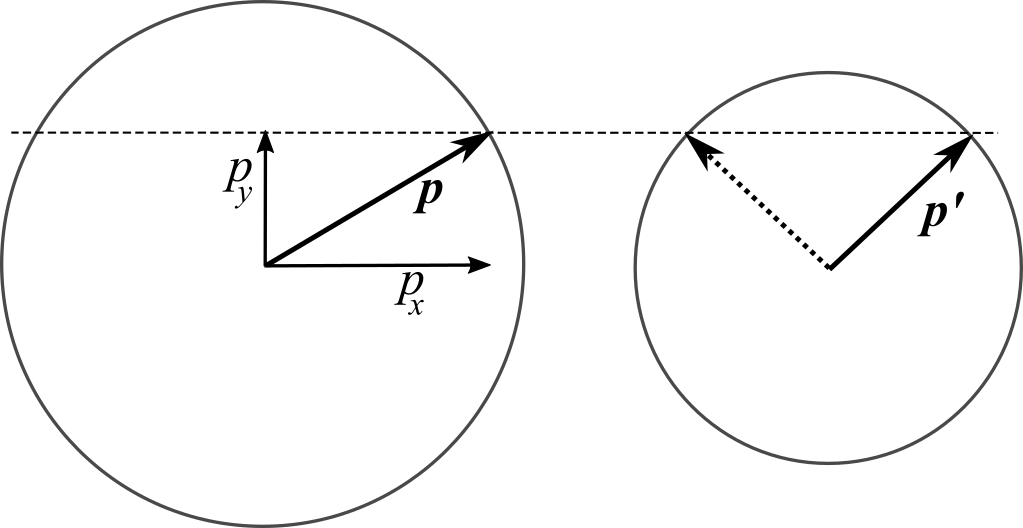}
\end{center}
\caption{Representation of the conservation of energy  $s c \abs{\pp} = s' c \abs{\pp'} + \deltaV,$ and $y$-momentum $p_y = p_y'$.
We assume that an electron with energy $E>0$ is scattered from the left to the right of the quantum strip. 
The two circles in the $\pp$-space correspond, respectively, to the sections $c\abs{\pp} = E$ and  $cs'\abs{\pp'} + \deltaV= E$ of the left and right 
cones, and the horizontal dashed line represents conservation of $p_y$. 
Assuming $\deltaV >0$, the continuous arrow at the right represents the scattered pseudomomentum $\pp'$ in the case $E>\deltaV$ ($s = +1$),
while the dashed arrow represents $\pp'$ in the case $0< E<\deltaV$ ($s = -1$).
Recall, in fact, that negative chirality is characterized by the pseudomeomentum being antiparallel to the direction of motion.}
\label{consfig} 
\end{figure}

It is here important to remark that if the left wave  (i.e.\ at $x<0$) is characterized by $(\pp,s)$ and the right wave  (i.e.\ at $x>0$) 
is characterized by $(\pp',s')$ (recall, however, that $p_y$ is conserved, meaning $p_y = p_y'$), then the parameters $p_x$, $s$, $p_x'$, $s'$ 
are related by the conservation of energy 
\BE
\label{CoE}
  s c \abs{\pp} = s' c \abs{\pp'} + \deltaV,
\EE  
as exemplified in Figure \ref{consfig}.
%
%
%

%
\subsection{Kinetic transmission conditions (KTC)}
\label{S3.1}
%
Solving the eigenvalue problem \eqref{SSE}  with the potential $V$ satisfying conditions {\bf H1} and {\bf H2} above, 
provides us with the scattering data,  i.e.\ the transmission and reflection coefficients.
For $i = 1,2$, we denote  by $T_s^i(\pp,s)$ and $R_s^i(\pp,s)$
the transmission and reflection coefficients from the left ($i = 1$) and from the right ($i = 2$).
They satisfy the following properties:
\begin{enumerate}
\item[{\bf P1.}]
\underline{unitarity}: $T_s^i(\pp) \geq 0$ and $R_s^i(\pp) \geq 0$, with $T_s^i(\pp) + R_s^i(\pp) = 1$;
\item[{\bf P2.}]
\underline{symmetry}: $T_s^i(\pp)$ and $R_s^i(\pp)$ are symmetric with respect to both $p_x$ and $p_y$;
\item[{\bf P3.}]
\underline{reciprocity}: $T_s^1(\pp) = T_{s'}^2(\pp')$, whenever $(\pp, s)$ and $(\pp',s')$ are related by the conservation of energy \eqref{CoE}.
\end{enumerate}
If the potential $V(x)$ is piecewise constant, as in some cases of importance for applications,  such as for the potential step \cite{CheianovEtAl2007}
or the potential barrier \cite{KatsnelsonEtAl06,Lejarreta13}, then the solution to \eqref{SSE} can be explicitly computed
by gluing up, with continuity, solutions of the form \eqref{planew}, procedure which allows to obtain explicit expressions of the scattering coefficients.
For example, for a potential step of height $\deltaV$  it is easy to check that the transmission coefficient for an electron incident from 
the left with energy $E = sc\abs{\pp}$, is given by
\BE 
\label{TTRR}
T_s^1(\pp)= \left\{ 
\begin{aligned}
&\frac{2 \cos(\phi) \cos(\theta)}{1 + \cos(\phi+\theta)},&\qquad &\text{if $\abs{E\sin(\phi)} < \abs{E-\deltaV}$,}
\\
&0,& &\text{otherwise,}
\end{aligned}
\right.
\EE
where  $\phi \in (-\frac{\pi}{2},\frac{\pi}{2})$ is the incidence angle and  $\theta \in (-\frac{\pi}{2},\frac{\pi}{2})$ is the transmission angle 
(both measured from an axis perpendicular to the step, so that  $\phi = 0$  and $\theta = 0$ correspond, respectively, to 
perpendicular incidence and transmission).
The two angles are constrained by
\BE 
\label{Snell}
E \sin(\phi) = (E-\deltaV) \sin(\theta). 
\EE
Note that \eqref{Snell} is a ``signed Snell law'': when $0 < E < \deltaV$ the angles $\phi$ and $\theta$ have opposite signs and 
$E-\deltaV$ is like a negative refractive index, which produces the electronic equivalent of the so-called Veselago lens \cite{CheianovEtAl2007}.
\par
\medskip
We now come to the kinetic part of the model, the quantum part being fully represented by the scattering reflection and transmission coefficients $T_s^i(\pp,s)$ and $R_s^i(\pp,s)$.
On the macroscopic scale, the quantum strip has a vanishing width and becomes a one-dimensional interface
between two classical regions.
Let us assume that, in addition to the quantum active potential $V(x)$ considered so far, there is a smooth potential $U(x,y)$, 
which can be neglected at the microscopic scale but becomes important in the classical regions  (where, conversely, $V$ is constant).
The kinetic description is expressed in terms of the phase-space distributions $w_s(\xx,\pp)$ of electrons with positive ($s=+$)
and negative ($s=-$) chirality/energy.
They are assumed to satisfy a semiclassical, stationary transport equation of the form
\BE
\label{WEw}
 \left( s\,\frac{c\pp}{\abs{\pp}}\cdot\nabla_\xx  - \nabla_\xx U\cdot\nabla_\pp\right) w_s = \cC_s(w_s),
\EE
where the left-hand side corresponds to the Hamiltonian dynamics \eqref{HSys} (with $V$ replaced by $U$) and  $\cC_s(w_s)$ is a suitable collisional term 
to be specified later on.
The semiclassical kinetic equation \eqref{WEw} is assumed to hold in the two classical regions, $x>0$ and $x<0$, for the two populations of electrons with
positive and negative chirality.
It is worth to remark that Eq.\ \eqref{WEw} has been introduced here in a heuristic way but it could be deduced as the semiclassical limit of the 
von Neumann (quantum Liouville) equation via Wigner transform \cite{Barletti14,ChapterKP}.
\par
Following Ref.\ \cite{NBA98}, we introduce a kinetic-quantum coupling in terms of 
{\em kinetic transmission conditions} (KTC) between the two classical regions through the quantum interface $x=0$.
The fundamental idea is that an incident/transmitted/reflected plane wave at the quantum interface, characterized by $(\pp, s)$, 
is identified with a corresponding particle in the classical regions inflowing/outflowing at $x=0$.
More precisely, since the direction of motion of an electron with pseudomomentum $\pp$ and chirality $s$
is $s\pp/\abs{\pp}$ (see Eqs.\ \eqref{HSys} and \eqref{sclvel}),
then such an electron is entering the left region (or leaving the right region) if $sp_x < 0$, and is leaving the left region 
(or entering the right region) if $sp_x > 0$.
\\
With this in mind, in order to express  the KTC, let us first of all introduce a suitable notation.
\begin{definition}
\label{notation}
An upper index $i = 1,2$ denotes the  left/right limits at $x=0$ of an $x$-dependent quantity $u(x)$:
$$
u^1 := \lim_{x \to 0^-} u(x),
\qquad\quad
u^2 := \lim_{x \to 0^+} u(x).
$$
\end{definition}
Then, according to what was discussed above, we write down the KTC as follows:
\BE
\label{TC0}
\left\{ 
\begin{aligned}
&w_s^1(p_x) = R_s^1(\pp) w_s^1(-p_x) + T_{s'}^2(\pp') w^2_{s'}(p_x'),&\quad &sp_x,\, s'p'_x < 0,
\\[8pt]
&w_{s'}^2(p'_x) = R_{s'}^2(\pp) w_{s'}^2(-p'_x) + T_s^1(\pp) w^1_s(p_x),& &s'p'_x,\, sp_x > 0,
\end{aligned}
\right.
\EE
where only the relevant variable $p_x$ has been explicitly indicated, and we recall that $(\pp,s)$ and $(\pp',s')$
are uniquely determined each other by \eqref{CoE} together with the indication of the sign of $p_x$ (this is enough, since $p_y = p_y'$).
In \eqref{TC0} we also used the fact that the reflection and transmission coefficients depend on $\pp$ and $\pp'$ only through 
$\abs{\pp}$ and $\abs{\pp'}$ (property 2 of the scattering coefficients).
\par
Recalling that the  products $sp_x$ and $s'p_x'$ determine the outflow or the inflow direction, it is easy to give the following interpretation
of the conditions \eqref{TC0}: at each side of the quantum interface, the inflow into the classical region
is given in part by the reflected outflow from the same side and in part by the transmitted outflow from the opposite side.
%
\subsection{Electrons and holes}
\label{S3.2}
%
%
The final goal of this work is to derive a diffusive limit of the hybrid kinetic model introduced in the previous section. 
This still needs a further step in the kinetic description, namely the specification of a suitable collision operator in Eq.\ \eqref{WEw},
and the consequent introduction of the hole population.
\par
To simplify the derivation of the diffusive equations we assume that the collisional term $\cC_s(w_s)$ is of Bhatnagar-Gross-Krook (BGK) type,
which expresses the relaxation of $w_s$ towards a local Fermi-Dirac (F-D) distribution having the same density as $w_s$.
Thus, we assume
\BE
\label{BGK}
\cC_s(w_s) = \frac{\weq_s - w_s}{\tau},
\EE
where 
\BE
\label{weq}
  \weq_s = \frac{1}{\e^{s(\beta c\abs{\pp}-A_s)}+1}.
\EE
Here, $\tau$ is the relaxation time and $\beta := 1/k_BT$, where $k_B$ is the Boltzmann constant and  $T$ is the given temperature.
Moreover,  the sign of the chemical potentials $A_s$ has been chosen for later convenience.%
\footnote{Note that here we are using non-dimensional chemical potentials, while the dimensional chemical potentials, that have the
dimensions of a energy, are given by $\beta^{-1}A_s$.}
\par
Now, the equilibrium $\weq_s$ should be related to the unknown distribution $w_s$ by the requirement  
that they have the same density.
However, since the lower energy cone is unbounded from below (see Eq.\ \eqref{cones}), 
$\weq_-$ cannot have finite moments, and such requirement does not make any sense for $s = -1$.
In order to fix this, we have to describe negative-energy/chirality electrons in term of electron vacancies
{\em (holes)}.
Let us therefore introduce the distributions $f_+$ (electrons) and $f_-$ (holes) defined by
\BE
\label{fdef}
  f_+(\xx,\pp) = w_+(\xx,\pp), \qquad
  f_-(\xx,\pp) =  1 - w_-(\xx,-\pp).
\EE
Note that the definition of $f_-$ contains a change in the sign of $\pp$, so that holes move parallel to $\pp$.
\par
By applying the transformation \eqref{fdef} to Eq.\ \eqref{WEw}, with $\cC_s$ given by \eqref{BGK}, we obtain
\BE
\label{WEfFD}
 \left(  \frac{c\pp}{\abs{\pp}}\cdot\nabla_\xx  - s \nabla_\xx U\cdot\nabla_\pp\right) f_s = 
\frac{\feq_s - f_s}{\tau},
\EE
where 
\BE
\label{feq}
\feq_s = \frac{1}{\e^{\beta c \abs{\pp}-A_s}+1}\,,
\EE
are now F-D distributions both with {\em positive} energies, so that they possess finite moments.
In particular, we can ask that $\feq_s$ and $f_s$ have the same densities,
i.e.\ we impose the constraint
\BE
\label{constf}
    \bk{\feq_s} = \bk{f_s} := n_s,
\EE
where we have introduced the bracket notation for the normalized%
\footnote{The normalization constant is required in order to get the the correct moments of a non-dimensional Wigner function \cite{Barletti14}.}
integrals
\BE
\label{bracket}
  \bk{\cdot} = \frac{1}{(2\pi\hbar)^2} \int_{\mR^2} \cdot\, d\pp.
\EE
Equation \eqref{WEfFD} with the constraint \eqref{constf}, is the stationary, semiclassical transport equation 
which shall be used for the description of electrons and holes transport in the semiclassical regions. 
\par
By using polar coordinates it is not difficult to see that the constraint \eqref{constf} 
fixes the chemical potentials $A_s$ as functions of the densities $n_s$, via the following formula (see \cite{Barletti14}):
\BE
\label{AvsN}
   \phi_2(A_s) = \frac{n_s}{n_0},
\EE
where
\BE
\label{phidef}
  \phi_k(z) := \frac{1}{\Gamma(k)} \int_0^\infty \frac{t^{k-1}}{e^{t-z} + 1}\,dt
\EE
and
\BE
\label{n0def}
  n_0 := \frac{2\pi}{(2\pi\hbar c\beta)^2}= \frac{(k_BT)^2}{2\pi \hbar^2 c^2}.
\EE
The function $\phi_k: \mR \to (0,+\infty)$ is the Fermi integral of order $k>0$, and can be
proved to be strictly increasing.
It will be convenient to denote by $A(n)$ the chemical potential corresponding to the density $n$, i.e.
\BE
\label{An}
   A(n) =  \phi_2^{-1}\left( \frac{n}{n_0}\right),
\EE
and to introduce the notation
\BE
\label{Fdef}
  F_n(\pp)  = \frac{1}{\e^{\beta c \abs{\pp}-A(n)}+1}\,,
\EE
for the F-D distribution with density $n$.
Then we shall rewrite the transport equation  \eqref{WEfFD} as
\BE
\label{SBE}
 \tau \left(  \vv\cdot\nabla_\xx  - s \nabla_\xx U\cdot\nabla_\pp\right) f_s = 
F_{\bk{f_s}} - f_s ,
\EE
which incorporates the constraint \eqref{constf} and where we adopted the notation 
\BE
\label{vdef}
 \vv(\pp) := \frac{c\pp}{\abs{\pp}}
\EE
for the semiclassical velocity.
Note that, at variance with negative-energy electrons, the velocity of holes has the same direction as $\pp$.
\subsection{Kinetic transmission conditions for electrons and holes}
\label{S3.3}
We now need to express the KTC \eqref{TC0} in terms of the distributions $f_s$, i.e.\ in terms of electrons and holes.
Let us begin by introducing more handy notations.
We define the ``state variable'' 
$$
  \zz = (\pp,s) = (p_x,p_y,s) \in \mR^2\times\{-1,+1\},
$$
and express all the quantities that depend on $\pp$ and $s$ as functions of $\zz$, e.g.\ the Dirac cones,
\BE
\label{cones}
 E(\zz) = E_s(\pp) =  sc\abs{\pp},
\EE
and the electron/hole densities
\BE
  f(\xx,\zz) = f_s(\xx,\pp).
\EE
Moreover, for $i = 1,2$ we define the following sets
\BE
\label{Phidef}
\begin{aligned}
  \Set &:= \mR^2 \times\{-1,+1\},
\\
  \Set^i_\IN &:= \{ \zz \in \Set \mid (-1)^i p_x >0 \},
\\
  \Set^i_\OUT &:= \{ \zz \in \Set \mid (-1)^i p_x < 0 \}.
\end{aligned}
\EE
Note that $ \Set^i_\IN$ and  $\Set^i_\OUT$ correspond, respectively, to the inflow and the outflow ranges of the pseudomomentum at $x = 0$, 
pertaining to the  left  ($i = 1$) and right ($i = 2$) regions.
The integration with respect to $\zz$ will stand for a sum with respect to $s$ and an integration with respect to $\pp$, 
for example:
$$
  \int_\Set f(\zz)\,d\zz = \sum_{s = \pm 1} \int_{\mR^2} f_s(\pp) d\pp.
$$
Instead, recall that $\bk{\cdot}$ (definition \eqref{bracket}) is just a normalized integration with respect to $\pp$ and, therefore, 
is a quantity that depends on $\xx$ and $s$:
\BE
\label{bknew}
   \bk{f}(\xx,s) = \frac{1}{(2\pi\hbar)^2} \int_{\mR^2} f(\xx,\zz) d\pp  = \frac{1}{(2\pi\hbar)^2} \int_{\mR^2} f_s(\xx,\pp) d\pp.
\EE
We also introduce the reflection transformation
\BE
  {\sim}\zz := (-p_x,p_y,s),
\EE
for $\zz = (\pp,s)$, which exchanges $\Set^i_\IN$ and  $\Set^i_\OUT$.
Note that the properties {\bf P1}--{\bf P3} of the scattering coefficients imply the following identities:
\begin{enumerate}
\item
$T^i(\zz) + R^i(\zz) = 1$;
\item
$T^i({\sim}\zz) = T^i(\zz)$ and $R^i({\sim}\zz) = R^i(\zz)$;
\item
$T^i(\zz) = T^j(\zz')$, if $E(\zz) = E(\zz') + (-1)^j\deltaV$, with $j\not= i$.
\end{enumerate}
\par
Now, by applying the transformation \eqref{fdef} to Eq.~\eqref{TC0}, and using the notation just introduced,
we can express the transmission conditions for the electron/hole distributions as follows:
\BE
\label{TC}
f^i(\zz) = R^i(\zz) f^i({\sim}\zz) + T^j(\zz') \left( ss' f^j(\zz') + \eps_{ss'}\right),\quad \zz \in \Set^i_\IN,\ \zz' \in \Set^j_\OUT ,
\EE
where $j \not= i$, and $\zz'$ is constrained to $\zz$ by the conservation of the energy and of the $y$-component of the pseudomomentum,
namely
\BE
\label{Econs}
  E(\zz) = E(\zz') + (-1)^ j\deltaV, \qquad p_y = p_y',
\EE
for $\zz = (\pp,s)$ and for $\zz' = (\pp',s')$.
The symbol $\eps_{ss'}$ is defined as
\BE
  \eps_{ss'} = \left\{
  \begin{aligned}
  &0, &\text{if $s = s'$},
  \\
  &1, &\text{if $s \not= s'$}.
  \end{aligned}
  \right.
\EE
The KTC in the form \eqref{TC} express very clearly the fact that the inflow at $\zz \in \Set^i_\IN$ is partly due to the reflected outflow 
${\sim}\zz \in \Set^i_\OUT$ from the same side $i$, and partly given by the transmitted outflow $\zz' \in \Set^j_\OUT$ from the opposite side $j$.
The inhomogeneous term $\eps_{ss'}$ comes from the inhomogeneous relation \eqref{fdef} between $f_s$ and $w_s$.
Moreover, Eqs.\ \eqref{TC} and \eqref{Econs}, together with the reciprocity property of the scattering coefficients, make the symmetry of the 
transmission conditions evident:
the equation for the $i$-side is is transformed in the equation for the $j$-side by changing the sign of $\deltaV$.
In particular, when $\deltaV$ = 0, the two equations are identical and (since in this case $\zz = \zz'$) take the simple form
\BE
\label{TC1.1}
f^i(\zz) = R^i(\zz) f^i({\sim}\zz) +  T^j(\zz) f^j(\zz),
\qquad \zz \in \Set^i_\IN.
\EE
\begin{proposition}[Flux conservation]
\label{Prop1}
For all $\deltaV \in \mR$ the KTC \eqref{TC} conserve the total charge flux across the quantum interface $x = 0$, i.e.
\BE
\label{Jcons}
  J^1_{+,x} - J^1_{-,x} = J^2_{+,x} - J^2_{-,x},
\EE
where, recalling definitions \eqref{bracket} and \eqref{vdef}, $\JJ_s = (J_{s,x},J_{s,y})$ is the current, defined by
\BE
\label{Jdef}
  \JJ_s := \bk{\vv f_s}.
\EE
If $\deltaV = 0$, then the conservation of the flux is valid separately for each population 
\BE
\label{Jcons2}
  J^1_{+,x} = J^2_{+,x},  \qquad  J^1_{-,x} = J^2_{-,x},
\EE
which means that there is no particle exchange between the upper and lower cone.
\end{proposition}
{\bf Proof.}
In order to incorporate more explicitly in the transmission conditions  the conservation properties \eqref{Econs}, let us rewrite \eqref{TC}  as follows:
\begin{multline}
\label{TC1}
f^i(\zz) = R^i(\zz) f^i({\sim}\zz) 
\\
+\int_{\Set^j_\OUT} T^j(\zz') \left( ss' f^j(\zz') + \eps_{ss'}\right)  k^j(\zz,\zz') \abs{\mu(\zz')} d\zz' ,
\quad \zz \in \Set^i_\IN,
\end{multline}
where $j \not= i$ and where we defined
\BE
\label{kdef}
  k^j(\zz,\zz') = \delta\big(E(\zz) - E(\zz')  - (-1)^j \deltaV\big)\,\delta(p_y - p_y'),
\EE
\BE
\label{mudef}
  \mu(\zz) = v_x(\pp)  = \frac{cp_x}{\abs{\pp}}.
\EE
Note that $\abs{\mu(\zz)}$ is the Jacobian determinant of the transformation
$$
  \zz \mapsto (E(\zz),p_y),
$$
which is bijective from $\Set_\IN^i$ (or $\Set_\OUT^i$) to $\{ (E, p_y) \in \mR^2 \mid \abs{E} \geq c\abs{p_y} \}$.
By using $R^i = 1 - T^i$, we can also rewrite \eqref{TC1} in the following form:
\begin{multline}
\label{TC2}
f^i(\zz) - f^i({\sim}\zz) =  -T^i(\zz) f^i({\sim}\zz) 
\\
+ \int_{\Set^j_\OUT} T^j(\zz') \left( ss' f^j(\zz') + \eps_{ss'}\right)  k^j(\zz,\zz') \abs{\mu(\zz')} d\zz' ,
\quad z \in \Set^i_\IN.
\end{multline}
Let us multiply both sides by $s\mu(\zz)$ and integrate over $\zz \in \Set^i_\IN$. 
At the left-hand side we obtain
$$
\int_{\Set^i_\IN} s\left[ f^i(\zz) - f^i({\sim}\zz)\right] \mu(\zz)\,d\zz
=
\int_{\Set^i_\IN}s f^i(\zz) \mu(\zz)\,d\zz + \int_{\Set^i_\OUT} sf^i(\zz) \mu(\zz)\,d\zz
$$ $$
 = \int_{\Set} s f^i(\zz) \mu(\zz)\,d\zz
 = \int_{\mR^2} f^i_+(\pp) \frac{cp_x}{\abs{\pp}}\,d\pp -  \int_{\mR^2} f^i_-(\pp) \frac{cp_x}{\abs{\pp}}\,d\pp,
$$ 
which is equal to $J^i_{+,x} - J^i_{-,x}$ upon multiplying by $(2\pi\hbar)^{-2}$.
At the right-hand side  we obtain
\begin{multline}
\label{aux7}
 - \int_{\Set^i_\IN} T^i(\zz) s f^i({\sim}\zz) \mu(\zz) d\zz 
 \\
 + \int_{\Set^i_\IN}  \int_{\Set^j_\OUT} T^j(\zz') \left( s' f^j(\zz') + s\eps_{ss'}\right)  k^j(\zz,\zz') \abs{\mu(\zz')}\mu(\zz) d\zz' d\zz
\\
=  \int_{\Set^i_\OUT} T^i(\zz)s f^i(\zz) \mu(\zz) d\zz +  
 \int_{\Set^j_\OUT}  T^j(\zz') s' f^j(\zz') \mu(\zz') d\zz' + C_j ,
\end{multline}
where we used the fact that $\abs{\mu(\zz')}\mu(\zz) = \mu(\zz')\abs{\mu(\zz)}$ for if $\zz \in \Set^i_\IN$ and $\zz' \in \Set^j_\OUT$,
and the identity
\BE
  \int_{\Set^i_\IN} k^j(\zz,\zz') \abs{\mu(\zz)} d\zz = 1.
\EE
The constant $C_j$ is%
\footnote{Note that the constant $C_i$ is finite because conservation of energy holds with different signs of $s$ and $s'$ only in a finite energy interval, 
which corresponds to a bounded region in $\pp$-space.}
\begin{multline*}
 C_j =  \int_{\Set^i_\IN}  \int_{\Set^j_\OUT} T^j(\zz') s\eps_{ss'}  k^j(\zz,\zz') \abs{\mu(\zz')}\mu(\zz) d\zz' d\zz
\\
  =  \int_{\Set^i_\OUT}  \int_{\Set^j_\IN} T^i(\zz) s'\eps_{s's}\,  k^i(\zz',\zz) \mu(\zz')\abs{\mu(\zz)} d\zz' d\zz = C_i \, ,
\end{multline*}
where we used the properties
$$
  k^j(\zz,\zz')  = k^i(\zz',\zz),  \qquad  s\eps_{ss'}  = - s'\eps_{s's}
$$
and made the change of variables $\zz \mapsto {\sim}\zz$, $\zz' \mapsto {\sim}\zz'$.
Hence, we see that the right-hand side expression \eqref{aux7} is identical for $(i,j) = (1,2)$ and $(i,j) = (2,1)$, which proves Eq.\ \eqref{Jcons}.
\\
In the particular case $\deltaV = 0$, the KTC reduce to the form \eqref{TC1.1} and, by rewriting them as
$$
f^i(\zz) - f^i({\sim}\zz) = -T^i(\zz) f^i({\sim}\zz) +  T^j(\zz) f^j(\zz),
$$
the verification of \eqref{Jcons2} is immediate.
$\square$

\begin{proposition}[KTC for Fermi-Dirac distributions]
\label{Prop2}
Let $n_+$ and $n_-$ be two assigned functions of $\xx$.
Then, the KTC \eqref{TC} (or their equivalent formulation \eqref{TC1})  are satisfied for the F-D distributions 
$$  
  f(\xx,\zz) = f_s(\xx,\pp) = F_{n_s(\xx)}(\pp), \qquad s = \pm,
$$ 
if and only if
\BE
\label{constrA}
  sA(n^1_s) = s'A(n^2_{s'}) + \beta \deltaV,
\EE
for all ``admissible'' couples $(s,s')$, i.e.\ such that 
\BE
\label{cons}
  sc\abs{\pp} = s'c\abs{\pp'} + \deltaV,
\EE
for some $\pp \not= 0$ and $\pp' \not= 0$ with  $T^1(\pp) \not= 0$.
\end{proposition}
{\bf Proof.}
We recall that \eqref{TC} are the KTC \eqref{TC0} after the transformation \eqref{fdef}. 
If $f_s(\xx,\pp) = F_{n_s(\xx)}(\pp)$, then the corresponding $w_s$'s are given by
$$
   w_s(\xx,\pp) =  \frac{1}{\e^{s\beta c\abs{\pp}-sA(n_s(\xx))}+1}.
$$
Substituting these F-D distributions in Eq.\ \eqref{TC0}  with $i=1$, using $R^1(\zz) = 1-T^1(\zz)$ and $T^1(\zz)  = T^2(\zz')$, and the fact that  
$w_s(\xx,\pp) = w_s(\xx,-\pp)$ (for such particular $w_s$'s), we obtain the condition
$$
  T^1_s(\pp) \left[w^1_s(\pp) - w^2_{s'}(\pp')\right] = 0
$$
(note that for $i= 2$ one would obtain to the same condition).
Hence, we find that the equality
$$
\frac{1}{\e^{s\beta c\abs{\pp}-sA(n^1_s)}+1} = \frac{1}{\e^{s'\beta c\abs{\pp'}-s'A(n^2_{s'})}+1}
$$
must hold for all $(\pp,s)$ and $(\pp',s')$ that satisfy \eqref{cons} with $T^1_s(\pp) \not= 0$.
This defines the admissible couples $(s,s')$, provided that $\pp \not= 0$ and $\pp'\not= 0$ (otherwise $s$ or $s'$
are undefined).
Substituting $s' c \abs{\pp'} = - s c \abs{\pp} + \deltaV$, we get the equality
$$
\frac{1}{\e^{s\beta c\abs{\pp}-sA(n^1_s)}+1} = \frac{1}{\e^{s\beta c\abs{\pp} - \beta\deltaV- s'A(n^2_{s'})}+1} ,
$$
which implies Eq.\ \eqref{constrA}.
$\square$

\begin{remark}
\label{remcouples}
It is readily seen that, apart from degenerate situations, the admissible couples $(s,s')$ are $(+,+), (+,-),(-,-)$ if $\deltaV > 0$; $(+,+), (-,-)$ if $\deltaV = 0$;
$(+,+), (-,+),(-,-)$, if $\deltaV < 0$. 
\end{remark}
%
%
\section{Diffusion limit}
\label{S4}
%
%
In this section we study the diffusion limit of the hybrid kinetic-quantum model \eqref{SBE}, \eqref{TC}, by assuming $\tau \ll 1$.
We divide the derivation, which is based on the Hilbert expansion method, into the ``bulk'' part (i.e. the semiclassical regions) and the
``interface'' part (i.e., close to the quantum interface).
\subsection{Diffusion limit in the semiclassical regions}
\label{S4.1}
Let us consider the Hilbert expansion (HE) of the unknown $f(\xx,\zz) = f_s(\xx,\pp)$ in \eqref{SBE}, in powers of the relaxation time $\tau$, to be considered as a small parameter:
\BE
\label{HE}
  f = f^{(0)} + \tau f^{(1)} +  \tau^2 f^{(2)} + \cdots
\EE
When substituting this expansion into the transport equation \eqref{SBE} we have to be aware of the fact that the BGK operator is nonlinear, due to the
use of F-D statistics $F_{\bk{f}}$.
A linearization of the collision operator is thus necessary, which requires to expand the F-D distribution around the equilibrium density, i.e.
\begin{multline}
\label{Fexp}
   F_{n^{(0)} + \tau n^{(1)} + \tau^2 n^{(2)}  + \cO(\tau^3)} = 
 \\  
   F_{n^{(0)}}  + \tau F'_{n^{(0)}} n^{(1)} 
   + \tau^2\big[F'_{n^{(0)}} n^{(2)} + \frac{1}{2}F''_{n^{(0)}}(n^{(1)})^2  \big]
   + \cO(\tau^3),
\end{multline}
where the primes denote the derivatives of $F_n$ with respect to $n$. 
By using \eqref{An} and the property $\phi_k' = \phi_{k-1}$, we obtain
\BE
\label{Fp}
  F'_n(\pp)  = \frac{ F_n(\pp)^2 \e^{\beta c \abs{\pp} - A(n)}}{n_0 \phi_1(A(n))},
\EE
while the explicit form of $F''_n$ is not important.
Moreover, note that
\BE
\label{bkFp}
  \bk{F'_n} = \frac{d}{dn}\bk{F_n} =  \frac{d}{dn}\,n = 1
\EE
and, for the same reason, one has
\BE
\label{bkFpp}
  \bk{F''_n} = 0.
\EE
The linearisation of our BGK collision operator
\BE
\label{Qdef}
  \cQ(f) := F_{\bk{f}} - f\,,
\EE
around the equilibrium $f^{(0)}$ is hence defined as
\BE \label{OPE}
  \cL_{\bk{f^{(0)}}}(g) := F'_{\bk{f^{(0)}}} \bk{g} - g.
\EE
In order to be able to find some information about the distribution functions $f^{(0)},  f^{(1)}, \ldots$, 
we shall need to study in more details this linear collision operator.
Note that $\cL_{\bk{f^{(0)}}}$ is an operator acting on functions of $\pp$, and the $(\xx,s)$-dependence is just
parametric, through the real parameter $n = \bk{f^{(0)}}(\xx,s)$.
The properties of $\cL_n$ are summarised in the following Lemma, whose proof is rather standard and 
can be easily adapted from \cite{Poupaud91}.
\begin{lemma}[Properties of the linearised collision operator $\cL_n$]
\label{Lprop}
Let $n \geq 0$ be a fixed real number, and let $\cL_n : \cH \rightarrow \cH$ be the operator defined 
by $\cL_n(g) := F'_n \bk{g} - g$ and acting on the Hilbert space
$\cH :=\rL^2\left( \mR^2, (F'_n)^{-1}(\pp) \,d\pp \right)$, with Hermitian product
$$
\quad (f,g)_\cH:= \int_{\mR^2} f(\pp)g(\pp)\, (F'_n)^{-1}(\pp)\, d\pp.
$$
\\
(i) $\cL_n$ is a well-defined, linear, bounded, symmetric and non-negative operator with kernel given by
$$
\ker \cL_n:= \left\{g \in \cH\: \mid \: g= F'_n \bk{g}\right\}\,.
$$
\\
(ii) The orthogonal of the kernel is nothing else than the range of $\cL_n$ and is given by
$$
(\ker \cL_n)^\perp = \mathrm{imag}\, \cL_n=\left\{g \in \cH \: \mid \:  \bk{g}= 0 \right\}\,.
$$
\\
(iii) Coercivity: for any $f \in \mathrm{imag}\, \cL_n$,
$$
  -  \int_{\mR^2}  \cL_n(f)\, f\, (F'_n)^{-1}\, d\pp \geq  \norma{f}_\cH^2\, .
$$
\\
(iv) Invertibility: the operator $\cL_n$ is a one-to-one mapping, if defined as
$$
\cL_n: (\ker \cL_n)^\perp \rightarrow (\ker \cL_n)^\perp\,,
$$
such that the equation $\cL_n(f)=g$ has a unique solution $f \in (\ker \cL_n)^\perp$ if and only if $g \in (\ker \cL_n)^\perp$.
\end{lemma}
Plugging now the HE \eqref{HE} into Eq.\ \eqref{SBE}, one obtains, at second order in $\tau$,
\begin{multline}
\label{HEE}
 \tau  \left( \vv\cdot\nabla_\xx  - s \nabla_\xx U\cdot\nabla_\pp \right) \left(f^{(0)} + \tau  f^{(1)} \right)= 
 \cQ(f^{(0)})  + \tau  \cL_{\bk{f^{(0)}}}(f^{(1)}) 
  \\[4pt]
 + \tau^2  \cL_{\bk{f^{(0)}}}(f^{(2)}) + \frac{\tau^2 }{2} F''_{\bk{f^{(0)}}} \bk{f^{(1)}}^2  + \cO(\tau^3)
\end{multline}
(we recall that $f^{(k)}$ and $\bk{f^{(k)}}$ are $s$-dependent quantities).
Comparing the terms of the same power in $\tau$ permits to get step by step some information on $f^{(0)}$, $ f^{(1)}$,\ldots, and finally to obtain the Drift-Diffusion model in the limit $\tau \rightarrow 0$.
\\[6pt]
{\underline {Step 1: order $\tau^0$.}} At order $\tau^0$ we obtain the condition $\cQ(f_s^{(0)})=0$ meaning $F_{\bk{f_s^{(0)}}}  = f_s^{(0)}$, which implies that the equilibrium
is a F-D distribution function
\BE
\label{tau0}
   f^{(0)}(\xx,\zz) =  f_s^{(0)}(\xx,\pp)  = F_ {n_s(\xx)}(\pp),
\EE
with $n_+(\xx)$ and $n_-(\xx)$ still to be determined.
\\[6pt]
\underline{Step 2: Order $\tau^1$.}
 At order $\tau^1$ we obtain the equation
$$
  \cL_{\bk{f^{(0)}}}(f^{(1)})   =  \left( \vv\cdot\nabla_\xx  - s \nabla_\xx U\cdot\nabla_\pp \right) f^{(0)}.
$$
By indicating the $s$-dependence explicitly and using \eqref{Fp} and \eqref{tau0}, this equation
can be rewritten as
\BE
\label{eqOrd1}
\cL_{n_s}(f_s^{(1)}) 
 = F'_{n_s} \vv \cdot \left(\nabla n_s + s\beta n_0\, \phi_1(A_s) \nabla U  \right)  =: -\eell_s \cdot \jj_s,
\EE
where, for later convenience, we have denoted
\BE
\label{LLdef}
   \eell_s = \left(\ell_{s,x}\, ,\ell_{s,y}\right)  := \frac{2}{c^2} F'_{n_s} \vv
\EE
and
\BE
\label{JJdef}
   \jj_s = \left(j_{s,x}\, ,j_{s,y}\right) := - \frac{c^2}{2} \left(\nabla n_s + s\beta n_0\, \phi_1(A_s) \nabla U \right).
\EE
Owing to the properties of the linearised BGK operator (see Lemma \ref{Lprop}), Eq.\ \eqref{eqOrd1} has the general solution
\BE
\label{tau1}
   f_s^{(1)}  =  \eell_s\cdot\jj_s + \sigma_s F'_{n_s},
\EE
for any $\sigma_s$ constant with respect to $\pp$.
Without loss of generality we can take $\sigma_s = 0$, meaning that $f_s^{(1)} \in (\ker \cL_{n_s})^\perp$,
because the addition of $\sigma_s F'_{n_s}$ does not affect the subsequent steps.
By using the the identity
$$
    \bk{\vv\otimes\vv F_n} = \frac{c^2 n}{2} I\,,
$$
where $I$ is the $2\times 2$ identity matrix (see e.g.\ Ref.\ \cite{Barletti14}), one obtains
$$
  \bk{\vv\otimes\vv F'_n}  =  \frac{d}{dn}\bk{\vv\otimes\vv F_n}  = \frac{c^2 }{2} I\,,
$$
leading to
\BE
\label{vell}
  \bk{\vv \otimes \eell_s} = I.
\EE
Equation \eqref{vell}, together with the obvious identities $\bk{\vv F_{n_s}} = \bk{\vv F'_{n_s}} = 0$, 
has the important implication that the current $\JJ_s=\bk{\vv\, f_s}$ is given by $\tau\jj_s$ up to higher orders, namely
\BE
\label{jvsJ}
  \JJ_s = \tau \jj_s+ \cO(\tau^2).
\EE
\\
\underline{Step 3: Order $\tau^2$.} 
Going on with the HE \eqref{HEE}, at order $\tau^2$ we get the equation
$$
  \cL_{\bk{f^{(0)}}}(f^{(2)})   = F'_{\bk{f^{(0)}}} \bk{f^{(2)}}  - f^{(2)}  
  = \left( \vv\cdot\nabla_\xx  - s \nabla_\xx U\cdot\nabla_\pp \right) f^{(1)} - \frac{1}{2} F''_{\bk{f^{(0)}}} \bk{f^{(1)}}^2.
$$
The solvability condition of this equation with respect to $f^{(2)}$ is
\begin{multline*}
  \bk{\left( \vv\cdot\nabla_\xx  - s \nabla_\xx U\cdot\nabla_\pp \right)  f^{(1)} - \frac{1}{2} F''_{\bk{f^{(0)}}} \bk{f^{(1)}}^2} =
\\
  \bk{\left( \vv\cdot\nabla_\xx  - s \nabla_\xx U\cdot\nabla_\pp \right)  f^{(1)}} = \bk{\vv\cdot\nabla_\xx f^{(1)}} = 0,
\end{multline*}
where the first equality comes from \eqref{bkFpp}.
This solvability condition is nothing else than the stationary
Drift-Diffusion equation for graphene \cite{Barletti14,MMR17},  whose explicit form, thanks to \eqref{vell}, is  readily found to be
\BE
\label{SDD1}
 \DIV\left[\nabla n_s + s\beta n_0\, \phi_1(A_s) \nabla U \right] = 0.
\EE
Note that $n_0\, \phi_1(A_s)$ is a nonlinear function of the
density $n_s$, which reduces to $n_s$  in the Maxwell-Boltzmann approximation (see Sec.\ \ref{S4.4}).
\\[6pt]
The results of this section are summarised in the following Proposition.
\begin{proposition}
\label{PropHE}
Supposing the solution $f(\xx,\zz) = f_s(\xx,\pp)$ of the transport equation \eqref{SBE} to admit a Hilbert expansion
of the form
\BE
\label{HE1}
  f = F + \tau G + \cO(\tau^2),
\EE
then the highest order terms are given by 
\BE
\label{FGdef}
   F(\xx,\zz)  = F_ {n_s(\xx)}(\pp) \qquad \text{and}  \qquad G(\xx,\zz)  = \eell_s(\xx,\pp)\cdot \jj_s(\xx),
\EE
where the functions  $F_ {n_s(\xx)}(\pp)$, $\ell_s(\xx,\pp)$ and  $\jj_s(\xx)$ are given, respectively, by \eqref{Fdef}, \eqref{LLdef} and \eqref{JJdef}.
Moreover, up to terms of order $\tau^2$, the density $n_s(\xx)$ satisfies the Drift-Diffusion equation \eqref{SDD1} in the semiclassical 
regions $x > 0$ and $x<0$.
\end{proposition}
%
\subsection{Diffusion limit at the quantum interface: leading order}
\label{S4.2}
%
We have now to deal with the diffusion limit of the transmission conditions. 
This means that we have to perform the Hilbert expansion of the left and right boundary values of $f(\xx,\zz)$, i.e.\ of $f^1(\zz)$ and $f^2(\zz)$.
\\
To this aim, let us introduce a concise notation for the Transmission Conditions \eqref{TC1}.
We put
\BE
\label{indef}
   f^i_{\IN} = f^i_{|\Set^i_\IN}, \qquad f^i_{\OUT} = f^i_{|\Set^i_\OUT}
\EE
(see definition \eqref{Phidef}), and rewrite the KTC \eqref{TC1} in a short form as
\BE
\label{TC3}
  f^i_{\IN} = \cB^i(f^i_{\OUT},f^j_{\OUT}),
\EE
where the boundary operator $\cB^i$ is defined as 
\BE
\cB^i(f^i_\OUT,f^j_\OUT)(\zz) =   R^i(\zz) f_\OUT^i({\sim}\zz) + T^j(\zz') \left( ss' f_\OUT^j(\zz') + \eps_{ss'}\right),
\quad 
 \zz \in \Set^i_\IN,
\EE
with $\zz' \in \Set^j_\OUT$ defined by \eqref{Econs} and with the usual convention that $j \not= i$.
It is important to remark that $\cB^i$ is not a linear but rather an affine transformation, so that
\BE
\label{BandK}
  \cB^i\left((f+g)^i_\OUT,(f+g)^j_\OUT\right) = \cB^i(f^i_\OUT,f^j_\OUT) + \cK^i(g^i_\OUT,g^j_\OUT),
\EE
the linear part $\cK^i$ being, of course,
\BE
\label{Kdef}
\cK^i(f^i_\OUT,f^j_\OUT)(\zz) =  R^i(\zz) f_\OUT^i({\sim}\zz) + T^j(\zz') f_\OUT^j(\zz').
\EE
We recall that the first two terms of the Hilbert Expansion $f = F + \tau G + o(\tau)$, far from the interface, 
are given in Proposition \ref{PropHE}.
Therefore, the KTC at leading order are
$$
  F^i_\IN = \cB^i \big(F^i_\OUT,F^j_\OUT \big),
$$
where $F(\xx,\zz)$ is the Fermi-Dirac distribution $F_ {n_s(\xx)}(\pp)$.
But, then, Proposition \ref{Prop2} applies, and leads to the following result.
\begin{proposition}[Diffusive transmission conditions (DTC) at leading order]
Up to terms of order $\tau$, the left and right densities at the interface $x = 0$, $n^1_s$ and $n^2_s$, are constrained by
the condition
\BE
\label{AA}
    sA(n^1_s) = s'A(n^2_{s'}) + \beta \deltaV,
\EE
that must hold for all admissible couples $(s,s')$.
\end{proposition}
The leading-order DTC \eqref{AA} are not ``quantum'', to the extent that they do not depend on the scattering coefficients.
In the next section we shall see that the first-order correction introduces such dependence.
%
%
\subsection{Diffusion limit at the quantum interface: first order}
\label{S4.3}
%
In order to search for a quantum correction, we require that the kinetic transmission conditions are satisfied by $f = F + \tau G + o(\tau)$
also at the first order in $\tau$ and, therefore, we impose the condition
\BE
\label{TC1tau}
  F^i_\IN + \tau G^i_\IN =  \cB^i\big(F^i_\OUT + \tau G^i_\OUT,F^j_\OUT + \tau G^j_\OUT \big).
\EE
However here we step into a difficulty, since, while 
$F^i_\IN =  \cB^i\big(F^i_\OUT,F^j_\OUT\big)$ can be satisfied with the suitable choice \eqref{AA} of the chemical potentials,
in general no chemical potentials exists such that Eq.\ \eqref{TC1tau} is also satisfied.
This means that the HE ansatz is incorrect at order  $\tau$ in the proximity of the interface. 
This difficulty is not new and similar situations are considered in literature. In general, one can overcome this burden by introducing a suitable boundary layer corrector \cite{DEA02,DS98}, which will lead to the well-known Milne-problems, permitting finally to couple the two Drift-Diffusion models on both sides of the interface.
\par
Let us present in more details how to obtain these Milne problems. Instead of using the Hilbert expansion \eqref{HE}, 
we shall slightly modify it by inserting a layer corrector at the order $\tau$, in the following manner
\BE
\label{HE2}
 f(x,y,\zz) = F(x,y,\zz) + \tau \left[ G(x,y,\zz) - H\Big({x\over \tau},y,\zz\Big) \right]  + \tau^2 f^{(2)}(x,y,\zz) + \cdots ,
\EE
where $F(\xx,\zz)$ and $G(\xx,\zz)$ are still given by Eq.\ \eqref{FGdef}, and $H(\xi,y,\zz)$ is the corrector term, 
which is a function of the boundary-layer variable by $\xi=x/\tau$. 
The corrector is to be chosen in such a manner to satisfy the following two requirements:
\begin{enumerate}
\item[\bf R1.]
the corrector should not affect the HE in the bulk, i.e.\ $F + \tau G - \tau H$, should be still a solution to the transport equation \eqref{SBE}
up to $\cO(\tau^2)$ far from the interface;
\item[\bf R2.]
at the interface, the corrector has to be constructed such that the transmission conditions at first-order are satisfied for a suitable
choice of chemical potentials.
\end{enumerate}
Substituting now the modified Hilbert-Ansatz \eqref{HE2} into Eq.\ \eqref{SBE} and denoting, for simplicity reasons, the transport term by 
$\cT:=\vv\cdot\nabla_\xx  - s \nabla_\xx U\cdot\nabla_\pp$, yields
\BE
\label{auxaux}
\tau \cT (F + \tau G ) - \tau \mu\, \frac{\pt H}{\pt \xi} =\cQ(F)  + \tau  \cL_{\bk{F}}(G)  - \tau \cL_{\bk{F}}(H) +  {\mathcal O}(\tau^2)\,,
\EE
where $\mu$, $\cQ$ and $\cL$ are defined in \eqref{mudef}, \eqref{Qdef} and \eqref{OPE}, respectively.
Using now the identities satisfied by $F$ and $G$, i.e.\ $Q(F) = 0$ and $\cT F = \cL_{\bk{F}}(G)$,  one remains, up to $\cO(\tau^2)$, with the equation
$$
\mu\, \frac{\pt H}{\pt \xi}  =  F'_{\bk{F}} \bk{H} - H.
$$
Recalling that 
$$
\bk{F}(\xx,s) = n_s(x,y) = n_s(\tau\xi,y),
$$ 
we expect to introduce just an error of order $\tau^2$ in \eqref{auxaux}, if we substitute  $n_s(x,y)$ with its boundary values $n^i_s(y)$, 
where we recall that  $n_s^i(y)$ denotes the limit of $n_s(x,y)$ as $x$ tends to $0$ from the $i$-th side (Definition \ref{notation}).
Thus, we expect that requirement {\bf R1} is fulfilled if $H$ satisfies the equation
\BE
\label{aux1}
\mu\, \frac{\pt H}{\pt \xi}  =  F'_{n_s^i} \bk{H} - H
\EE
on the $i$-th side of the interface (this is nothing else than a half-space, stationary transport equation).
We shall see below that a corrector satisfying \eqref{aux1} and such that also requirement {\bf R2} is satisfied,  
can be constructed by means of four auxiliary functions
$$
 \theta^i(\xi,y,\zz) =  \theta_s^i(\xi,y,\pp), \quad (-1)^i \xi>0, \quad  y\in\mR, \quad \zz = (\pp,s) \in \Set, \quad i = 1,2,
$$
that satisfy \eqref{aux1} associated with the linear non-homogeneous KTC
$$
  \theta^i_\IN - G^i_\IN =  \cK^i\big(\theta^i_\OUT - G^i_\OUT,\theta^j_\OUT - G^j_\OUT \big),
$$
(recall \eqref{BandK} and \eqref{Kdef}).
Hence, let us consider the problem 
\BE
\label{Milne}
\left\{
\begin{aligned}
&\mu \frac{\pt \theta^i}{\pt \xi}  = F'_{n^i_s} \bk{\theta^i} - \theta^i,&  (-1)^i &\xi>0,
\\[4pt]
&\theta^i_\IN - \cK^i\big(\theta^i_\OUT,\theta^j_\OUT \big) =  G^i_\IN - \cK^i\big(G^i_\OUT,G^j_\OUT \big),&  &\xi=0,
\end{aligned}
\right.
\EE
Note that in $n^i_s$, the upper index $i$ refers to the interface limit of $n_s$, but in $\theta^i$ it is used to label the side where the problem 
is posed (and not the limit).
Note also that the coordinate $y$ is just an overall parameter in the problem.
\par
Equation \eqref{Milne} is a system of four half-space, half-range Milne problems \cite{DM79}, for the functions $\theta^1_+$, $\theta^1_-$,
$\theta^2_+$, $\theta^2_-$, coupled via non-homogenous transmission conditions.
The following theorem, whose proof is deferred to  Appendix \ref{appe}, is fundamental for the construction of the layer corrector. 
\begin{theorem}[Solution to the coupled Milne problems]
\label{T1}
Problem \eqref{Milne} admits a solution $(\theta^1,\theta^2)$ with 
$$
\theta_s^i \in \rL^\infty\big( (-1)^i[0,+\infty)\times \Set,  (F'_{n^i_s})^{-1}d\xi d\pp \big),
\qquad
i = 1,2, \quad s = \pm,
$$ 
if and only if the flux conservation condition holds:
\BE
\label{jcond}
\begin{cases}
  j^1_{+,x} - j^1_{-,x} = j^2_{+,x} - j^2_{-,x} ,& \text{if $\deltaV \not= 0,$}
  \\[4pt]
  j^1_{s,x}  = j^2_{s,x}, \quad s = \pm1,  & \text{if $\deltaV = 0.$}
\end{cases}
\EE
Such solution is unique up to the addition of any solution of
the homogeneous problem (i.e., problem \eqref{Milne} with $G = 0$).
Moreover, four constants $\as^{1,\infty}_+$, $\as^{1,\infty}_-$, $\as^{2,\infty}_+$, $\as^{2,\infty}_-$ 
(depending on the parameter $y$) exist such that
\BE
\label{asymptheta}
    \theta^i(\xi,y,\zz) \to  \theta^{i,\infty}(y,\zz) := \as_s^{i,\infty}(y) F'_{n^i_s} (\pp), \quad \text{as $\xi \to (-1)^i\infty$,}
\EE
and the convergence is exponentially fast; 
in particular, 
\BE
\label{damping}
\abs{\bk{\theta_s^i} -  \as_s^{i,\infty}} \leq C\e^{-\alpha\abs{\xi}},
\EE
for some constants $C>0$ and $\alpha>0$ (possibly depending on the parameter $y$).
\end{theorem}
Thanks to Theorem \ref{T1} we can now construct the corrector $H$, which we define as follows:
\BE
\label{ChiDef}
H(\xi,y,\zz)  = 
\left\{
\begin{aligned}
&\theta^1\left(\xi,y,\zz \right) -  \theta^{1,\infty}(y,\zz),& &\text{if $\xi < 0$,}
\\[4pt]
&\theta^2\left(\xi,y,\zz \right) -  \theta^{2,\infty}(y,\zz),& &\text{if $\xi > 0$,}
\end{aligned}
\right.
\EE
where the functions $\theta^i$, are the solution to the coupled Milne problem \eqref{Milne} 
and $\theta^{i,\infty}$ are their asymptotic distributions \eqref{asymptheta}.
As we shall see below,  although the solution to \eqref{Milne} is only  determined up to the addition of an arbitrary solution of the homogeneous problem, 
such addition does not affect the final result, namely Theorem \ref{T2}.
\\
Let us now verify that the corrector function $H$ given by \eqref{ChiDef} satisfies the two requirements {\bf R1} and {\bf R2}.
\\
First of all, from Theorem \ref{T1} it follows immediately that  $H$ vanishes exponentially fast away from the interface,  with 
\BE
\label{Hdamping}
\abs{\bk{H}} \leq C \e^{-\alpha\abs{x}/\tau}.
\EE
Moreover, $H$ satisfies Eq.\ \eqref{aux1}, since both $\theta_s^i$ and $\as_s^{i,\infty}F'_{n^i_s}$ do. 
Then, we already know that $F + \tau(G-H)$ satisfies the transport equation up to terms of order $\tau^2$ 
if the error that is made by substituting in \eqref{auxaux} $F'_{n_s} \bk{H}$ with the boundary limit $F'_{n^i_s} \bk{H}$ is of order $\tau$.
But, indeed, from the Taylor expansion of $n_s$  (assumed regular enough) and from inequality \eqref{Hdamping}, we have that
$$
 \abs{( F'_{n^i_s} - F'_{n_s}) \bk{H}} \leq M \abs{x} \e^{-\alpha\abs{x}/\tau}  \leq \frac{M \tau}{\alpha} \e^{-\alpha\abs{x}/\tau} ,
$$
for some constant $M>0$.
This proves {\bf R1}.
\par
Coming to requirement {\bf R2}, when evaluating the  transmission conditions on the modified HE \eqref{HE2}, 
Eq.\ \eqref{TC1tau} is replaced by
\begin{multline}
\label{TC1tau2}
  F^i_\IN + \tau (G-H)^i_\IN =  \cB^i\left(F^i_\OUT + \tau (G-H)^i_\OUT,F^j_\OUT + \tau (G-H)^j_\OUT \right)
\\[4pt]
  =  \cB^i\big(F^i_\OUT ,F^j_\OUT\big) + \tau \cK^i\left((G-H)^i_\OUT,(G-H)^j_\OUT \right),
\end{multline}
where we used \eqref{BandK}.
But 
$$
(G-H)^i_\IN  - \cK^i\left((G-H)^i_\OUT,(G-H)^j_\OUT \right)
= \theta^{i,\infty}_\IN - \cK^i (\theta^{i,\infty}_\OUT,  \theta^{j,\infty}_\OUT),
$$
because of the definition of $H$ \eqref{ChiDef} and the boundary conditions in \eqref{Milne}.
Then, \eqref{TC1tau2} becomes 
\BE
\label{bc1st}
  F^i_\IN + \tau \theta^{i,\infty}_\IN  =   \cB^i\big(F^i_\OUT+\tau \theta^{i,\infty}_\OUT ,F^j_\OUT+\tau\theta^{j,\infty}_\OUT\big).
\EE
Now, we recall that $F^i_s = F_{n^i_s}$, $\theta^{i,\infty}_s = \as^{i,\infty}_s F'_{n^i_s}$,  and, from \eqref{Fexp}, 
$$
   F_{n_s^i} + \tau \as_s^{i,\infty} F'_{n_i^s} = F_{n^i_s + \tau \as_s^{i,\infty}} + \cO(\tau^2).
$$
Hence, Eq.\ \eqref{bc1st} is, up to $\cO(\tau^2)$, a KTC condition for Fermi-Dirac distributions with densities $n^i_s + \tau \as_s^{i,\infty}$
and Proposition \ref{Prop2} immediately leads to our principal result.
\begin{theorem}[Diffusive transmission conditions at first order]
\label{T2}
Assume that  the current conservation \eqref{jcond} holds. 
Then, up to errors of order $\tau^2$, the left and right densities at the interface $x = 0$, $n^1_s$ and $n^2_s$, are constrained by
the condition
\BE
\label{AA2}
    sA(n^1_s + \tau \as_s^{1,\infty}) = s'A(n^2_{s'}+ \tau \as_{s'}^{2,\infty}) + \beta \deltaV,
\EE
that must hold for all admissible couples $(s,s')$,
where  $\as_{s}^{i,\infty}$ are the asymptotic densities%
\footnote{Since we are using dimensionless phase-space distributions, the physical dimensions of $\as_s^{i,\infty}$ are actually those of a frequency.}
of the solution $\theta^i_s$ to the Milne problem \eqref{Milne}  (see \eqref{asymptheta}).
\end{theorem}
Equation \eqref{AA2} gives the DTC at first order for the  electron/hole densities $n_s$ across the interface.
They are a first order correction to the leading order conditions \eqref{AA} and can be considered as a  ``quantum correction''
since they depend upon the scattering data $T_s^i$ through the asymptotic densities  $\as_{s}^{i,\infty}$
associated to the solutions of the Milne problem \eqref{Milne}.
\begin{remark}
The solutions $\theta_s^i$ and, consequently, the asymptotic densities $\as_{s}^{i,\infty}$ are unique
only up to the addition of a solution to the homogeneous Milne problem.
However, since the addition of such a  solution does not change Eq.\ \eqref{bc1st} (by definition), 
then Eq.\ \eqref{AA2} is not affected by the particular choice of the solutions $\theta_s^i$. 
\end{remark}
\subsection{Maxwell-Boltzmann approximation}
\label{S4.4}
For large energies, the F-D distribution \eqref{Fdef}  is asymptotically approximated by the Maxwell-Boltzmann (M-B) distribution $n M(\pp)$, where
\BE
\label{Mdef}
M(\pp) =  \frac{\e^{-\beta c \abs{\pp}} }{n_0} 
\EE
is the normalised Maxwellian and the constant $n_0$ is given by \eqref{n0def}.
Correspondingly, the Fermi integrals \eqref{phidef} are asymptotically approximated by 
$$
\phi_k(A) \sim \e^A
$$
(independently on $k>0$) and, in particular, Eq.\ \eqref{An} is approximated by
$$
  \e^{A(n)} \sim \frac{n}{n_0}.
$$
Then, it is readily seen that the M-B approximation of the Drift-Diffusion equation \eqref{SDD1} is
given by
\BE
\label{SDD2}
  \DIV \left[\nabla n_s + s \beta n_s \nabla U\right] = 0,
\EE
and the M-B approximation of the first-order DTC \eqref{AA2} writes
 \BE
\label{MBDTC}
    \Big(\frac{n^1_s + \tau \as_s^{1,\infty}}{n_0}\Big)^s = \Big(\frac{n^2_{s'}+ \tau \as_{s'}^{2,\infty}}{n_0}\Big)^{s'} \e^{\beta \deltaV},
\EE
for all admissible couples $(s,s')$.
Moreover, the asymptotic densities $\as_{s}^{i,\infty}$ are calculated, as functions of $\jj_s^i$, from the Milne problem \eqref{Milne} with
\BE
\label{MilneMB}
F'_{n_s^i} \sim M,
\qquad
\eell^i_s \sim \frac{2}{c^2} M \vv .
\EE
Note that in the M-B approximation the dependence of $F'_n$ (and, consequently, $\eell$) on $n$ disappears and then the approximated 
quantities become independent on the indices $s$ and $i$.
\\
It is instructive to write down \eqref{MBDTC} more explicitly for $\delta V > 0$. 
In this case, the admissible couples $(s,s')$ are $(+,+), (+,-), (-,-)$ and we obtain therefore:
\BE
\label{zones}
\left\{
\begin{aligned}
& n^1_+ -  \e^{\beta \deltaV} n^2_+ = \tau\big(  \e^{\beta \deltaV}\as_+^{2,\infty} - \as_+^{1,\infty} \big),
\\[6pt]
&\big( n^1_+ + \tau \as_+^{1,\infty}\big)\big( n^2_- + \tau \as_-^{2,\infty}\big) =  \e^{\beta \deltaV} (n_0)^2,
\\[6pt]
& n^1_- -  \e^{-\beta \deltaV} n^2_- = \tau\big(  \e^{-\beta \deltaV}\as_-^{2,\infty} - \as_-^{1,\infty} \big).
\end{aligned}
\right.
\EE
We note that the first equations is identical (in form) to the first-order DTC found in Refs.\ \cite{DEA02,DS98} 
for the case of a single, parabolic energy band, and the third one is its hole version (the potential changes sign).
The second equation is a quantum correction to the semiconductor mass-action law $n^1_+n^2_ - = \e^{\beta \deltaV} (n_0)^2$.
Clearly, it can be approximated at order $\tau$ as follows:
\BE
\label{zones2}
  n^1_+  n^2_- - e^{\beta \deltaV} (n_0)^2 = - \tau\big( n^1_+ \as_-^{2,\infty}  +  n^2_ - \as_+^{1,\infty} \big).
\EE

\section{Hybrid Drift-Diffusion-quantum model}
\label{S5}
%
We now summarize the results obtained in the present work, by writing down the hybrid diffusive-quantum model 
describing the electron transport in a graphene device.
\par
\begin{figure}[htbp] 
\begin{center}
\includegraphics[width=.6\linewidth]{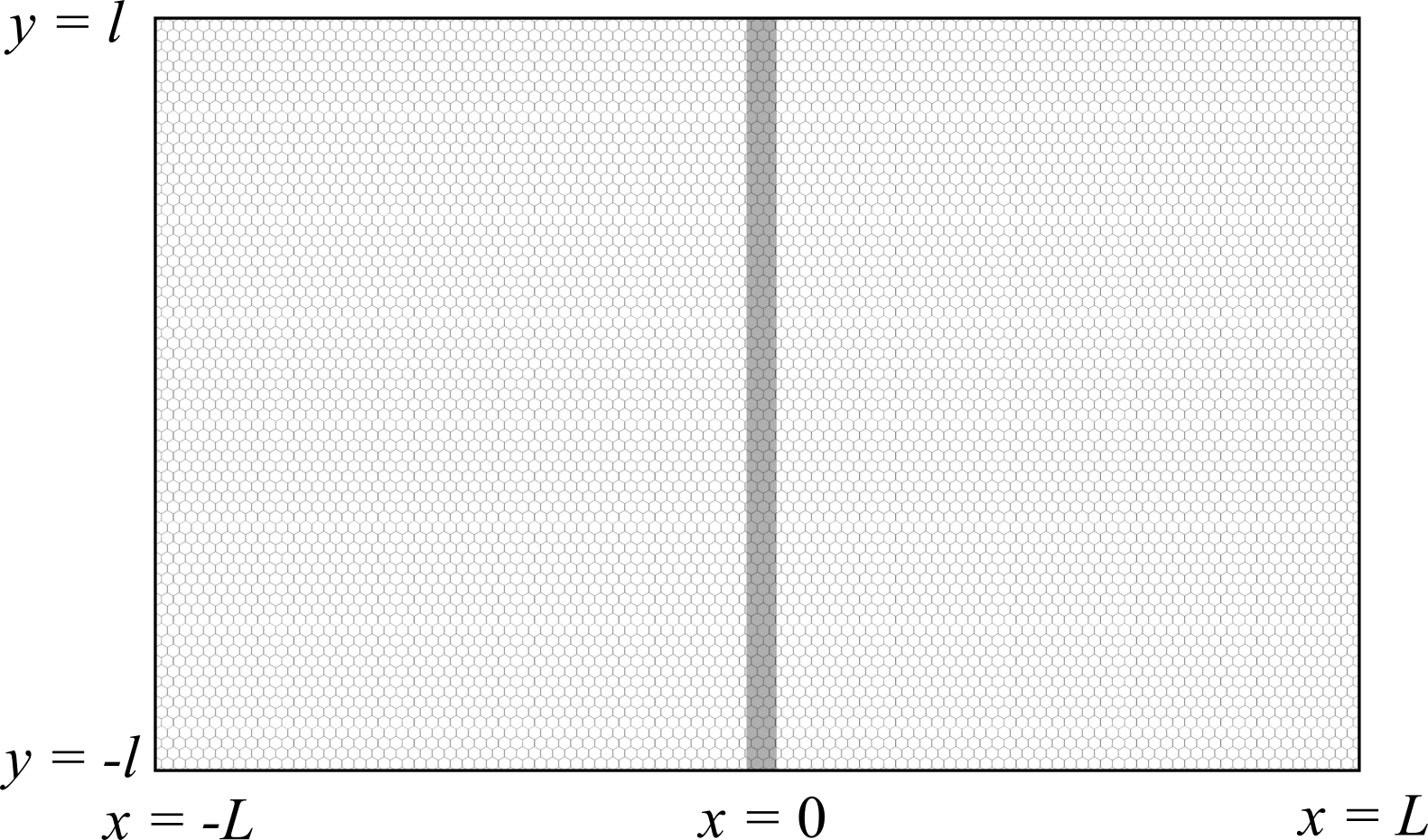}
\caption{Geometry of a prototypical graphene device: the central strip represents the quantum active region, where the DTC are imposed.
On the remaining boundaries, classical Dirichlet and Robin conditions can be imposed.}
\label{device}
\end{center}
\end{figure}

\smallskip
Let our hypothetic graphene device be represented by the rectangle
$(x,y) \in (-L,L) \times (-l, l)$ (see Fig.\ \ref{device}),
where the steep potential variations are concentrated in $x=0$ (on a macroscopic scale), and the two classical regions
are 
$$
   \Omega^1 = (-L,0) \times (-l, l),
   \qquad 
   \text{and}
   \qquad
   \Omega^2 = (0,L) \times (-l, l).
$$
For the sake of simplicity, we shall work in the M-B approximation (see Sec.\ \ref{S4.4}).
Then,  in $\Omega^1$ and $\Omega^2$ the stationary drift-diffusion equation 
\eqref{SDD2} is assumed to hold. 
At the external boundary of the device standard conditions can be imposed, e.g.\ non homogeneous Dirichlet
conditions at $x = -L$ and $x = L$ (representing ohmic contacts) and homogeneous Robin conditions at $y = -l$
and $y = l$ (representing an insulating boundary).
At the quantum-classical interface, $x=0$, the DTC \eqref{MBDTC} are imposed. 
\par
\smallskip
If $\deltaV > 0$, the DTC are explicitly given by \eqref{zones}, where the second equation can be substituted by \eqref{zones2}.
This is a rank-3 condition and we still need a further condition, which is given by the total flux 
conservation \eqref{jcond} across the interface. 
We stress the fact that the flux conservation is also required to ensure existence of  $\as_s^{i,\infty}$, according to  Theorem \ref{T1}.
\par
\smallskip
The resulting hybrid diffusive-quantum model reads as follows:
\\\
\begin{subequations}
\label{QDM}
\begin{itemize}
\item
in the semiclassical regions $\Omega^1\cup\Omega^2$:
\BE
\label{QDMa}
\left\{
  \begin{aligned}
  &\DIV \jj_s = 0,
  \\
  & \jj_s = -\big(\nabla n_s + s \beta n_s \nabla U\big),
  \end{aligned}
\right.
\EE
\item
at the Ohmic boundary $x = \pm L$:
\BE
\label{QDMb}
 n_s = n_s^{\pm L},
\EE
\item
at the insulating boundary $y = \pm l$:
\BE
\label{QDMc}
j_{s,y} = 0,
\EE
\item
across the quantum interface $x = 0$:
\BE
\label{QDMd}
 \left\{
\begin{aligned}
& n^1_+ -  \e^{\beta \deltaV} n^2_+ = \tau\big(  \e^{\beta \deltaV}\as_+^{2,\infty} - \as_+^{1,\infty} \big),
\\[4pt]
& n^1_+  n^2_- + \tau\big( n^1_+ \as_-^{2,\infty}  +  n^2_ - \as_+^{1,\infty} \big) = \e^{\beta \deltaV} (n_0)^2,
\\[4pt]
& n^1_- -  \e^{-\beta \deltaV} n^2_- = \tau\big(  \e^{-\beta \deltaV}\as_-^{2,\infty} - \as_-^{1,\infty} \big),
\\[4pt]
& j^1_{+,x} - j^1_{-,x} = j^2_{+,x} - j^2_{-,x}= 0.
\end{aligned}
\right.
\EE
\end{itemize}
\end{subequations}
In Eq.\ \eqref{QDMb}, $n_s^{\pm L}$ denote the given densities of electrons and holes at the contacts.
\par
The quantities $\as_s^{i,\infty}$ are the asymptotic densities associated to the system of Milne equations 
\BE
\label{Milne2}
\left\{
\begin{aligned}
&\mu\frac{\pt \theta_s^i}{\pt \xi}   = M \bk{\theta_s^i} - \theta_s^i,& \ (-1)^i \xi&>0,
\\[4pt]
&\theta^i_\IN - \cK^i\big(\theta^i_\OUT,\theta^j_\OUT \big) =  G^i_\IN - \cK^i\big(G^i_\OUT,G^j_\OUT \big),&  \xi&=0.
\end{aligned}
\right.
\EE
where
\BE
\label{GMB}
  G^i_s = M\vv\cdot\jj^i_s.
\EE
Note that the functions $\as_s^{i,\infty}(y)$ depend on $\jj_s$ and, therefore, the interface conditions \eqref{QDMd}
couple the four (left/right, electrons/holes) drift-diffusion equations \eqref{QDMa}.
We also remark that the $\as_s^{i,\infty}$'s embody the quantum part of the model, 
represented by the scattering problem \eqref{SSE}.
\begin{remark}
As it can be seen by comparing \eqref{GMB} with \eqref{MilneMB} and \eqref{FGdef}, here $\jj$ has been conveniently 
redefined as the drift-diffusion current divided by $c^2/2$.
\end{remark}
If $\deltaV = 0$, then the second DTC equations \eqref{zones} disappears and we are left with a rank-2 condition.
On the other hand, in this case, the current conservation holds {\em separately} for electrons and holes (see Eq.\  \eqref{jcond})
and we gain one more condition on the current. 
If $\deltaV = 0$, therefore, the diffusion model is  still given by \eqref{QDM} but the interface conditions \eqref{QDMd} must be substituted by 
\BE
 \left\{
\begin{aligned}
&n^1_+ - n^2_+ = \tau\big(  \as_+^{2,\infty} - \as_+^{1,\infty} \big),
\\[2pt]
&n^1_- -   n^2_- = \tau\big(  \as_-^{2,\infty} - \as_-^{1,\infty} \big),
\\[2pt]
&j^1_{+,x} = j^2_{+,x},
\\[2pt]
&j^1_{-,x} = j^2_{-,x},
\end{aligned}
\right.
\EE
We note that, in this case, also the asymptotic densities $ \as_s^{i,\infty}$ are decoupled with respect to $s$. 
This is an immediate consequence of the fact that, when $\deltaV = 0$, the KTC for electrons and holes are decoupled and, consequently, 
electrons and holes are also decoupled in the Milne problem \eqref{Milne}, or \eqref{Milne2} in the M-B case.
Hence, if no additional coupling mechanisms are introduced, if $\deltaV = 0$ electrons and holes are completely independent, both in the kinetic 
and in the diffusive models.
\par
\smallskip
Of course, solving numerically the coupled Milne equations \eqref{Milne} or \eqref{Milne2} is in general a hard task, 
and the advantage of the diffusive-quantum model \eqref{QDM}, with respect to the kinetic-quantum one, 
is far from being evident.
It is therefore necessary to reduce the complexity of problem \eqref{Milne}.
This can be done by assuming that the outflow distribution $\theta^i_\OUT$ is an equilibrium distribution (e.g.\ a Maxwellian, for
problem \eqref{Milne2}), so that the only unknowns of the albedo problems are four albedo densities (see Ref.\  \cite{DEA02} for 
the case of standard particles). 
Another possibility is the application of the iterative procedure proposed by Golse and Klar in Ref.\ \cite{GK95}.
This will be the subject of a subsequent work, devoted to numerics for real applications. 
\par
\smallskip
We finally remark that, in the present formulation, the quantum part of the problem is independent of the semiclassical one, 
to the extent that the scattering problem \eqref{SSE} is solved, self-consistently and once for all, in order to get 
the scattering data. 
However, a more complicate nonlinear coupling can be introduced by assuming that the quantum potential $V$ depends 
in part on the densities $n_s$ through a Poisson equation \cite{NBA98,DEA02}.
\section*{Acknowledgements}
Support is acknowledged from the Italian-French project PICS (Projet International de Coop\'eration Scientifique) 
``MANUS - Modelling and Numerics for Spintronics and Graphene'' (Ref. PICS07373).

\appendix

\section{Proof of Theorem \ref{T1}}
\label{appe}
In order to streamline the notation, let us put 
\BE
\label{Ldef}
  L^i(\zz) = L^i_s(\pp) := F'_{n^i_s}(\pp),
\EE
and rewrite the Milne problem \eqref{Milne} accordingly:
\BE
\label{MilneL}
\left\{
\begin{aligned}
&\mu \frac{\pt \theta^i}{\pt \xi}  = L^i \bk{\theta^i} - \theta^i,&  (-1)^i &\xi>0,
\\[4pt]
&\theta^i_\IN - \cK^i\big(\theta^i_\OUT,\theta^j_\OUT \big) =  G^i_\IN - \cK^i\big(G^i_\OUT,G^j_\OUT \big),&  &\xi=0.
\end{aligned}
\right.
\EE
We recall that in problem \eqref{MilneL} the $y$-variable is just a parameter, which shall be omitted throughout the proof.
\\
The proof is inspired by the ideas of Ref.\ \cite{DS98} and is divided into three steps for the reader's convenience.
\\[4pt]
{\underline{Step 1: reduction to a $\abs{\pp}$-averaged problem.}}
Let us consider the uncoupled version of \eqref{MilneL}, with assigned inflows $g^i$:
\BE
\label{UCM}
\left\{
\begin{aligned}
&\mu \frac{\pt \theta^i}{\pt \xi}(\xi,\zz)   = L^i (\zz)\bk{\theta^i} - \theta^i(\xi,\zz),&\quad (-1)^i &\xi>0,\quad  \zz \in \Set,
\\[4pt]
&\theta^i_\IN(\zz) =  g^i(\zz),&  &\xi=0, \quad \zz \in \Set^i_\IN
\end{aligned}
\right.
\EE
(recall definitions \eqref{Ldef}, \eqref{bknew} and \eqref{indef}).
We introduce the $\abs{\pp}$-average 
\BE
 \tilde\theta^i(\xi,\vphi,s) := \frac{1}{2\pi\hbar^2} \int_0^{+\infty}  \theta^i(\xi,\abs{\pp}\cos\vphi,\abs{\pp}\sin \vphi,s)\,\abs{\pp}\,d\abs{\pp},
\EE
where the normalisation constant is chosen so that
$$
  \bk{\theta^i}(\xi,s)  = \frac{1}{2\pi}\int_0^{2\pi} \tilde\theta^i(\xi,\vphi,s)  \,d\vphi 
  \quad \text{and} \quad
  \tilde L^i_s = 1,
$$
where $ \bk{\theta^i}$ is defined in \eqref{bracket}.
We introduce the analogous of the sets \eqref{Phidef} for the averaged quantities:
\BE
\label{PhiTdef}
  \tilde\Set := [0,2\pi) \times\{-1,+1\},
  \quad
  \tilde\Set^i_{\IN/\OUT} := \{ (\vphi,s) \in \Set \mid (-1)^i \cos\vphi \gtrless 0 \}
\EE
and extend, in the obvious way, to $\tilde\theta^i$ the notations $\tilde\theta^i_\IN$ and  $\tilde\theta^i_\OUT$.
Taking the $\abs{\pp}$-average of \eqref{UCM}, we obtain that $\tilde\theta^i_s$ satisfies  
\BE
\label{avM}
\left\{
\begin{aligned}
&\mu \frac{\pt \tilde\theta^i}{\pt \xi}(\xi,\vphi,s)  = \frac{1}{2\pi} \int_0^{2\pi} \tilde\theta^i(\xi,\vphi,s)\,d\vphi  - \tilde\theta^i(\xi,\vphi,s),& \ 
(-1)^i &\xi>0,\ (\vphi,s) \in \tilde \Set,
\\[4pt]
&\tilde \theta^i_\IN(\vphi,s) =  \tilde g^i(\vphi,s),&  &\xi=0,\ (\vphi,s) \in \tilde \Set^i_\IN ,
\end{aligned}
\right.
\EE
where we recall that 
$$
  \mu =  \frac{cp_x}{\abs{\pp}} = c\cos\vphi.
$$
Conversely, it is easy to check that, if $\tilde\theta^i$ is a solution of \eqref{avM}, then
\BE
\label{extension}
   \theta^i := L^i\tilde \theta^i + \left\{
   \begin{aligned}
    &\e^{-\xi/\mu}\left(  g^i -  L^i \tilde g^i \right),& &(-1)^i \mu>0,
    \\
    &0,& &(-1)^i \mu<0,
    \end{aligned}
   \right.
\EE
is solution of \eqref{UCM}.
The equations \eqref{avM} are four ($s = \pm1$, $i = 1,2$) independent Milne problems having the form of a 
Milne problem for neutron transport, to the extent that the kernel of the collision operator coincides 
with the functions that are constant with respect to $\vphi$ \cite{BSS84,DM79}.
About such problem the following facts are known \cite{BSS84,DS98,Poupaud91}:
\begin{itemize}
\item[(i)]
If $\tilde g^i \in \rL^\infty(\tilde \Set^i_\IN)$, the solution $\tilde \theta^i$ to problem  \eqref{avM} exists and is unique in
$\rL^\infty\big( (-1)^i[0,+\infty)\times \tilde \Set \big)$.
Moreover, one has the positivity, i.e.\ $\tilde\theta^i \geq 0$ if $\tilde g^i \geq 0$.
\item[(ii)]
A constant $\as_s^{i,\infty}$ (depending on $\tilde g^i$) exists such that $\tilde\theta^i(\xi,\vphi,s) \to \as_s^{i,\infty}$, as $\xi \to (-1)^i\infty$,
and the convergence is exponentially fast; in particular
$$
  \Big\vert \int_0^{2\pi} \tilde\theta^i(\xi,\vphi,s)\, d\vphi -  \as_s^{i,\infty} \Big\vert  \leq C\e^{-\alpha\abs{\xi}},
$$
for some constants $C>0$ and $\alpha>0$.
Moreover, $\as^{i,\infty} \geq 0$ if $\tilde g^i \geq 0$.
\item[(iii)]
The {\em Albedo operator}, that associates the inflow to the outflow, i.e.\  $ \tilde \theta^i_\IN \equiv \tilde g^i \mapsto \tilde \theta^i_\OUT$,
is a compact linear operator from $\rL^\infty(\tilde\Set^i_\IN)$ to $\rL^\infty(\tilde\Set^i_\OUT)$.
\end{itemize}
{\underline{Step 2: formulation as a Fredholm problem.}}
Thanks to the explicit formula \eqref{extension}, it is easy to extend the above results to the $\abs{\pp}$-dependent problem \eqref{UCM}.
Let us define the weighted spaces
$$  
 X^i :=  \rL^\infty\big( \Set,  (L^i)^{-1} d\pp \big),  
 \  
 X^i_\IN :=  \rL^\infty\big(\Set^i_\IN,  (L^i)^{-1} d\pp \big),
 \
 X^i_\OUT :=  \rL^\infty\big(\Set^i_\OUT,  (L^i)^{-1} d\pp \big).
$$
Note that $g \in  X^i$ implies that $g \in  \rL^p(\Set)$ for all $p \in [0,\infty]$.
Then, from the results of Step 1, we have the following facts about problem \eqref{UCM}:
\begin{itemize}
\item[(i)]
If $g^i \in X^i_\IN$, the solution $\theta^i$ to problem \eqref{UCM} exists and is unique in the space
$\rL^\infty\left((-1)^i[0,+\infty), X^i \right)$.
Moreover, $\theta^i \geq 0$ if $g^i \geq 0$.
\item[(ii)]
A constant $\as_s^{i,\infty}$ (depending on $g^i$) exists such that $\theta^i(\xi,\zz) \to \as_s^{i,\infty} L_s^i(\pp)$, as $\xi \to (-1)^i\infty$,
and the convergence is exponentially fast; in particular
$$
   \Big\vert \bk{\theta_s^i}(\xi) -  \as_s^{i,\infty} \Big\vert  \leq C\e^{-\alpha\abs{\xi}},
$$
for some constants $C>0$ and $\alpha>0$.
Moreover, $\as^{i,\infty} \geq 0$ if $g^i \geq 0$.
\item[(iii)]
The Albedo operator, associating the inflow $\theta^i_\IN \equiv g^i$ to the outflow $\theta^i_\OUT$,
$$
  \cA^i : X^i_\IN \to X^i_\OUT, \qquad \cA^i \theta^i_\IN := \theta^i_\OUT,
$$
is a compact linear operator.
\end{itemize}
We now come to the coupled Milne problem \eqref{MilneL}. 
Thanks to the Albedo operator just introduced, we can reformulate \eqref{MilneL} as a Fredholm problem
in $X^1_\IN\times X^2_\IN$ for the unknown inflow data $(\theta^1_\IN,\theta^2_\IN)$, namely:
\BE
\label{Fredholm}
   \begin{pmatrix} \theta^1_\IN \\[2pt]  \theta^2_\IN \end{pmatrix}
   -\cK  \begin{pmatrix} \cA^1 \theta^1_\IN \\[2pt]  \cA^2 \theta^2_\IN \end{pmatrix}
   =   \begin{pmatrix} \Gamma^1\\[2pt]  \Gamma^2 \end{pmatrix}
\EE
where, recalling definition \eqref{Kdef},
$$
  \cK  \begin{pmatrix} \theta^1_\OUT \\  \theta^2_\OUT \end{pmatrix} =
\begin{pmatrix}  \cK^1\big(\theta^1_\OUT, \theta^2_\OUT \big)  
\\[2pt]
\cK^2\big(\theta^2_\OUT, \theta^1_\OUT \big)  \end{pmatrix},
$$
and the components of the non-homogeneous term are
\BE
\label{Gammadef}
 \Gamma^i =  G^i_\IN - \cK^i\big(G^i_\OUT,G^j_\OUT \big).
\EE
Let us now show that $\cK : X^1_\OUT\times X^2_\OUT \to X^1_\IN\times X^2_\IN$
is a linear, continuous operator.
In fact, for $\zz \in \Set^i_\IN$ we can write 
 $$
 \frac{\theta^i_\IN(\zz)}{L^i(\zz)} =  \frac{\cK^i(\theta^i_{\OUT},\theta^j_{\OUT})(\zz)}{L^i(\zz)}
 = R^i(\zz) \frac{\theta^i_\OUT({\sim}\zz)}{L^i(\zz)} + T^j(\zz')  \frac{ss'\theta^j_\OUT(\zz')}{L^i(\zz)},
 $$
 where $\zz' \in \Set^j_\OUT$ is constrained to $\zz$ by the conservation laws \eqref{Econs}.
 Recalling definitions \eqref{Fp} and \eqref{Ldef}, and using \eqref{AA} and the identity
$$
  \frac{\e^h}{(\e^h+1)^2} =  \frac{\e^{-h}}{(\e^{-h}+1)^2},
$$
it is not difficult to show that the following relation holds
\BE
\label{Maux1.2}
  \phi_1(A(n^i_s))\, L^i_s(\pp) = \phi_1(A(n^j_{s'})) \, L_{s'}^j(\pp'),
\EE
for all $\zz = (\pp,s)$ and $\zz' = (\pp',s')$ related as above. 
Then, the previous equality can be rewritten as 
$$
 \frac{\theta^i_\IN(\zz)}{L^i(\zz)} =
 R^i(\zz) \frac{\theta^i_\OUT({\sim}\zz)}{L^i(\zz)} + T^j(\zz')  \frac{ss' c^i_s \theta^j_\OUT(\zz')}{c^j_{s'} L^j(\zz')},
 $$
 where
\BE
\label{cdef}
  c^i(\zz) =  c^i_s := \phi_1(A(n^i_s))
\EE
are positive constants that only depend on $s$ (and not on $\pp$).
Using Jensen inequality we can write
 $$
 \Big\vert \frac{\theta^i_\IN(\zz)}{L^i(\zz)} \Big\vert^2 \leq
 R^i(\zz)  \Big\vert \frac{\theta^i_\OUT({\sim}\zz)}{L^i(\zz)}  \Big\vert^2 
 + T^j(\zz')  \Big\vert \frac{c^i_s}{c^j_{s'}}  \Big\vert^2 \Big\vert \frac{\theta^j_\OUT(\zz')}{ L^j(\zz')} \Big\vert^2,
 $$
(we adopt the redundant notation $\abs{\cdot}^2$ for the square, just to improve readability),
 which shows the continuity of $\cK$, since the scattering coefficients are bounded by 1.
 \par
 Hence, the composition of $\cK$, which is continuous, with the $\cA^i$'s, which are compact, is a compact operator on
  $X^1_\IN\times X^2_\IN$ and, therefore, \eqref{Fredholm} is a Fredholm equation with compact operator.
 The proof of Theorem \ref{T1} is thus reduced to a Fredholm alternative, which will be discussed in the next two steps.
 \\[4pt]
{\underline{Step 3: the homogeneous problem.}}
We now consider the homogeneous version of the Milne problem \eqref{MilneL}, corresponding to $G=0$:
\BE
\label{Milne0}
\left\{
\begin{aligned}
&\mu \frac{\pt \theta^i}{\pt \xi} = L^i \bk{\theta^i}  - \theta^i ,&      (-1)^i \xi>0,
\\[4pt]
&\theta^i_\IN - \cK^i\big(\theta^i_\OUT,\theta^j_\OUT \big) =  0,&  \xi=0.
\end{aligned}
\right.
\EE
%
Let $(\theta^1,\theta^2)$ be a solution of such a problem in the space $X^1\times X^2$. 
It is convenient to introduce the functions $(\psi^1,\psi^2)$ as follows:
$$
    \theta^i(\xi,\zz) = \sqrt{L^i(\zz)} \, \psi^i(\xi,\zz), \qquad i = 1,2,
$$
so that $\psi^i/\sqrt{L^i}$ is bounded and $\psi^i$ satisfies the equation
\BE
\label{psieq}
  \mu \frac{\pt \psi^i}{\pt \xi}  = \sqrt{L^i} \, \bk{\sqrt{L^i} \, \psi^i} - \psi^i.
\EE
It is immediate to verify that 
\BE
\label{inequaux}
  \int_{\mR^2} \left(\sqrt{L_s^i}\, \Big\langle\sqrt{L_s^i}\, \psi_s^i\Big\rangle - \psi_s^i\right) \psi^i \, d\pp \leq 0,
\EE
and that the equality holds if and only if $\psi_s^i$ is in the kernel of  the collision operator, i.e.\  $\psi_s^i = \sqrt{L_s^i} \, \gamma_s$,
for some $\gamma_s$ constant with respect to $\pp$.
By multiplying by $\psi^i$ both sides of \eqref{psieq}, integrating over $\pp \in \mR^2$ we obtain
\BE
\label{Maux0}
  \frac{1}{2} \frac{\pt}{\pt \xi} \int_{\mR^2}  \abs{\psi_s^i(\xi,\pp)}^2  \mu(\pp)\, d\pp
  =  \int_{\mR^2} \left(\sqrt{L_s^i}\, \Big\langle\sqrt{L_s^i}\, \psi_s^i \Big\rangle- \psi_s^i\right) \psi_s^i \, d\pp
  \leq 0.
\EE
From (ii) of Step 2 we know that $\psi_s^i(\xi,\pp)$, as $\xi \to (-1)^i\infty$, tends to a function of the form $\sqrt{L_s^i(\pp)}\, \gamma_s^i(\xi)$ and, 
therefore, by integrating the previous inequality over $\xi \in (-1)^i[0,+\infty)$ and recalling that $\mu$ is an odd function of $\pp$, we obtain
\BE
\label{Maux1}
 \int_{\mR^2} \abs{\psi_s^1(0,\pp)}^2  \mu(\pp) \, d\pp \leq 0 \leq  \int_{\mR^2} \abs{\psi_s^2(0,\pp)}^2  \mu(\pp) \, d\pp.
\EE
On the other hand, $(\theta^1,\theta^2)$ satisfy the homogeneous KTC
$$
\theta^i(0,\zz) = R^i(\zz) \theta^i(0,{\sim}\zz) + T^j(\zz') ss' \theta^j(0,\zz')
$$
and, correspondingly,  $(\psi^1,\psi^2)$ satisfy
$$
 \sqrt{L^i(\zz)}\, \psi^i(0,\zz) = R^i(\zz) \sqrt{L^i(\zz)}\, \psi^i(0,{\sim}\zz) + T^j(\zz') ss' \sqrt{L^j(\zz')}\, \psi^j(0,\zz'),
$$
where $\zz \in \Set^i_\IN$ and $\zz' \in \Set^j_\OUT$ are constrained by the conservation of energy \eqref{Econs}.
By using \eqref{Maux1.2} and \eqref{cdef}, we obtain 
\BE
\label{TCpsi}
   \sqrt{c^i(\zz)}\, \psi^i(0,\zz) = R^i(\zz) \sqrt{c^i(\zz)}\, \psi^i(0,{\sim}\zz) + T^j(\zz') ss' \sqrt{c^j(\zz')}\, \psi^j(0,\zz'),
\EE
and then, using Jensen inequality, 
$$
c^i(\zz) \abs{\psi^i(0,\zz)}^2 \leq R^i(\zz) c^i(\zz) \abs{\psi^i(0,{\sim}\zz)}^2 + T^j(\zz') c^j(\zz') \abs{\psi^j(0,\zz')}^2
$$
or, equivalently,
\begin{multline*}
c^i(\zz) \left( \abs{\psi^i(0,\zz)}^2- \abs{\psi^i(0,{\sim}\zz)}^2\right)
\\
 \leq - T^i(\zz) c^i(\zz) \abs{\psi^i(0,{\sim}\zz)}^2 + T^j(\zz') c^j(\zz')\abs{\psi^j(0,\zz')}^2.
\end{multline*}
We now multiply both sides by $\mu(\zz)$, $\zz \in \Set^i_\IN$, which is negative (or zero) for $i = 1$ and positive (or zero) for $i = 2$, 
so that
\begin{multline*}
c^1(\zz) \left( \abs{\psi^1(0,\zz)}^2- \abs{\psi^1(0,{\sim}\zz)}^2\right) \mu(\zz) \geq 
- T^1(\zz) c^1(\zz) \abs{\psi^1(0,{\sim}\zz)}^2\mu(\zz)  
\\
  + T^2(\zz') c^2(\zz')\abs{\psi^2(0,\zz')}^2\mu(\zz),
\end{multline*}
\begin{multline*}
c^2(\zz) \left( \abs{\psi^2(0,\zz)}^2- \abs{\psi^2(0,{\sim}\zz)}^2\right) \mu(\zz) \leq 
- T^2(\zz) c^2(\zz) \abs{\psi^2(0,{\sim}\zz)}^2\mu(\zz)  
\\
  + T^1(\zz') c^1(\zz')\abs{\psi^1(0,\zz')}^2\mu(\zz).
\end{multline*}
If we now integrate the first inequality over $\zz \in \Set^1_\IN$, and the second one over $\zz \in \Set^2_\IN$, 
by following the same passages as in the proof of Proposition \ref{Prop1} we arrive at
\begin{multline*}
\int_\Set c^1(\zz) \abs{\psi^1(0,\zz)}^2 \mu(\zz) d\zz  \geq  
\int_{\Set^1_\OUT} T^1(\zz) c^1(\zz) \abs{\psi^1(0,\zz)}^2 \mu(\zz)\,d\zz
\\
+ \int_{\Set^2_\OUT} T^2(\zz') c^2(\zz') \abs{\psi^2(0,\zz')}^2 \mu(\zz')\,d\zz'
\end{multline*}
\begin{multline*}
\int_\Set c^2(\zz) \abs{\psi^2(0,\zz)}^2 \mu(\zz) d\zz   \leq 
 \int_{\Set^2_\OUT}  T^2(\zz') c^2(\zz')  \abs{\psi^2(0,\zz')}^2 \mu(\zz')\,d\zz'
\\
+ \int_{\Set^1_\OUT}  T^1(\zz) c^1(\zz)  \abs{\theta^1(0,\zz)}^2 \mu(\zz)\,d\zz
\end{multline*}
(where in the last integral of both inequalities, $\zz$ and $\zz'$ are constrained by $E(\zz) = E(\zz') + \deltaV$),
which immediately leads to
\BE
\label{Maux2}
\int_\Set c^1(\zz) \abs{\psi^1(0,\zz)}^2 \mu(\zz) d\zz  \geq  
\int_\Set c^2(\zz) \abs{\psi^2(0,\zz)}^2 \mu(\zz) d\zz.
\EE
%
If we now come back to \eqref{Maux1}, multiply the right and the left sides by the positive constants $c^1(\zz) = c^1_s$
and $c^2(\zz) = c^2_s$, respectively, and sum up with respect to $s$, we obtain 
\BE
\label{Maux2.1}
\int_\Set c^1(\zz) \abs{\psi^1(0,\zz)}^2 \mu(\zz) d\zz  \leq 0 \leq   
\int_\Set c^2(\zz) \abs{\psi^2(0,\zz)}^2 \mu(\zz) d\zz.
\EE
By comparing \eqref{Maux2} with \eqref{Maux2.1} we see that, necessarily,
$$
 \int_\Set c^1(\zz) \abs{\psi^i(0,\zz)}^2 \mu(\zz) d\zz = 0, \qquad i = 1,2.
$$
Multiplying  \eqref{Maux0} by $c^i(\zz) = c^i_s$, summing up with respect to $s$ and integrating with respect to $\xi$ yields, therefore, 
$$
  \int_0^{(-1)^i\infty} \int_\Set c^i \left(\sqrt{L^i}\, \Big\langle\sqrt{L^i}\, \psi^i \Big\rangle- \psi^i\right) \psi^i \, d\zz\,d\xi= 0.
$$
Since $c^i$ are positive constants that only depend on $s$, and the integrals with respect to $\pp$ are definite in sign (see \eqref{inequaux}), 
this implies that
$$
   \int_{\mR^2} \left(\sqrt{L_s^i}\, \Big\langle\sqrt{L_s^i}\, \psi_s^i \Big\rangle- \psi_s^i\right) \psi_s^i \,d\pp = 0
$$
for all $\xi \in (-1)^i[0,+\infty)$.
This equality can only hold when $\psi^i(\xi,\cdot)$ is in the kernel of the collision operator, which implies that $\theta^i(\xi,\pp)$ is necessarily
of the form
$$
  \theta^i(\xi,\zz) = \theta_s^i(\xi,\pp) = L_s^i(\pp) \gamma_s^i(\xi).
$$
Finally, the substitution of this expression in the first of equations \eqref{Milne0} immediately yields that $\gamma_s^i$ is constant, so that 
\BE
\label{Maux3}
  \theta_s^i(\xi,\pp) = L_s^i(\pp) \gamma_s^i.
\EE
Substituting \eqref{Maux3} in the second of equations \eqref{Milne0} and using \eqref{Maux1.2} leads to the following necessary and sufficient 
condition for \eqref{Maux3} to be solution of the homogeneous Milne problem \eqref{Milne0}:
\BE
\label{Maux4}
 c_{s'}^j\gamma_s^i = c_s^i \gamma_{s'}^j 
\EE
where $c^i_s$ is defined by \eqref{cdef}, and $i,j$ and $s,s'$ are related as usual.
Recalling that the conservation of energy is satisfied by three couples $(s,s')$ if $\deltaV \not= 0$ and just by two couples in the case if $\deltaV = 0$ 
(see Remark \ref{remcouples}), we notice that \eqref{Maux4} is a rank-3 condition if $\deltaV \not= 0$ and a rank-2 condition if $\deltaV = 0$.
 \\[4pt]
{\underline{Step 4: the inhomogeneous problem.}}
Let us finally return to the complete, inhomogeneous problem \eqref{MilneL}.
Let $(\theta^1,\theta^2)$ be a bounded solution of \eqref{MilneL}.
The integration in $\pp$ of the first equation in \eqref{MilneL} yields
$$
  \frac{d}{d\xi} \bk{\mu \theta_s^i} = 0,
$$
and the integration in $\pp$ after multiplication by $\mu$ yields
$$
   \frac{d}{d\xi} \bk{\mu^2 \theta_s^i} = - \bk{\mu \theta_s^i}.
$$
Hence, $\bk{\mu \theta_s^i}$ is a constant, and this constant must be zero, otherwise  $\abs{\bk{\mu^2 \theta_s^i}}$ 
would grow linearly with $\xi$, in contradiction with the boundedness assumption.
So we have
\BE
\label{thetaflux}
   \int_{\mR^2} \theta^i_s(\xi,\pp)\,\mu(\pp)\,d\pp = 0,
\EE
for all $\xi \in (-1)^i[0,+\infty)$, $i =1,2$ and $s = \pm1$.
If we now rewrite the boundary conditions as
$$
\theta^i_\IN - G^i_\IN - \cK^i\big(\theta^i_\OUT - G^i_\OUT,\theta^j_\OUT - G^j_\OUT \big) =  0,
$$
then, from Proposition \ref{Prop1} (that applies also to the linear KTC), we have that the conservation of charge flux holds:
$$
  \int_\Set s(\theta^1(0,\zz)- G^1(\zz)) \mu(z)\,d\zz = \int_\Set s(\theta^2(0,\zz)- G^2(\zz)) \mu(\zz)\,d\zz.
$$
But then, since \eqref{thetaflux} implies that the charge flux associated to $\theta^i$ vanishes, we obtain
that the flux conservation  for $G^i$ must hold:
\BE
\label{suffcond}
  \int_\Set s\,G^1(\zz) \mu(\zz)\,d\zz =  \int_\Set s\, G^2(\zz) \mu(\zz)\,d\zz.
\EE
Equation \eqref{suffcond} is therefore a necessary condition for the existence of a bounded solution to the Milne problem \eqref{MilneL} or, equivalently, 
to the Fredholm problem \eqref{Fredholm} when $\Gamma^i$ has the form \eqref{Gammadef}.
Using \eqref{vell}, it is immediate to verify that the $G^i$'s, given by \eqref{FGdef}, satisfy this condition if and only if \eqref{jcond} holds.
\\
Now, from Step 3 we know that the kernel of the Fredholm operator
at the left-hand side of \eqref{Fredholm} is spanned by the functions of the form $L^i_s(\pp)\gamma^i_s$, with $\gamma^i_s$ satisfying \eqref{Maux4}, 
and is therefore a subspace of  $X^1_\IN,\times X^2_\IN$ of dimension $d$, where $d = 1$ 
if $\deltaV \not = 0$, and $d = 2$ if $\deltaV = 0$.
Hence, the range of the Fredholm operator is a closed subspace of codimension $d$. 
But \eqref{suffcond} also defines a subspace of codimension $d$, and then it describes the condition of existence of the solution  
to the Fredholm equation when $\Gamma^i$ is of the form \eqref{Gammadef}.
We conclude that \eqref{jcond} is a necessary and sufficient condition for the existence of a solution to the Fredholm equation \eqref{Fredholm}
(and, therefore, of the Milne problem \eqref{MilneL}), up to a solution of the associated homogeneous problem. 
This proves the first part of Theorem \ref{T1}.
\\
The second part of the theorem, that is the existence of the asymptotic densities $\as_s^{i,\infty}$ and the exponential estimate \eqref{damping}, 
follows from (ii) of Step 2. 
In fact, once the coupled Milne problem is solved, the inflow of each component $\theta^i_s$ is determined (up to the addition of a term of the form
\eqref{Maux3}--\eqref{Maux4}), and point (ii) of Step 2 applies.%
\footnote{Except positivity, that is not guaranteed (and not required)  here.}


\begin{thebibliography}{}

\bibitem{Ashcroft}
Ashcroft, N.W., Mermin. N.D.: 
Solid State Physics.
Saunders College Publishing, Philadelphia (1976)

\bibitem{BSS84}
Bardos, C., Santos, R., Sentis, R.:
Diffusion approximation and the computation of the critical size.
T. Am. Math. Soc. 284, 617--649 (1984)

\bibitem{Barletti14}
Barletti, L.: .
Hydrodynamic equations for electrons in graphene obtained from the maximum entropy principle.
J. Math. Phys. 55, 083303 (2014)

\bibitem{Barletti16}
Barletti, L.: 
Hydrodynamic equations for an electron gas in graphene.
J. Math. Ind.  6:7 (2016)

\bibitem{BarlettiNegulescu17}
Barletti, L., Negulescu, C.:
Hybrid classical-quantum models for charge transport in graphene with sharp potentials.
 J. Comput. Theor. Transport 46, 159--175 (2017)

\bibitem{ChapterKP}
Barletti, L., Frosali, G., Morandi, O.:
Kinetic and hydrodynamic models for multi-band quantum transport in crystals.
In: Ehrhardt, M., Koprucki, T. (eds.)
Multi-band Effective Mass Approximations: Advanced Mathematical Models 
and Numerical Techniques, pp.\ 3--56. Springer, Heidelberg (2014)

\bibitem{NBA98}
Ben Abdallah, N.:
A hybrid kinetic-quantum model for stationary electron transport.
J. Stat. Phys. 90, 627--662  (1998)

\bibitem{NBA02}
Ben Abdallah N., Degond P., Gamba I.:
Coupling one-dimensional time-dependent classical and quantum transport models. 
J. Math. Phys.  43, 1--24 (2002)

\bibitem{Borysenko10}
Borysenko, K.M., Mullen, J.T.,  Barry, E.A. , Paul, S.,  Semenov, Y.G., Zavada, J.M. , Buongiorno Nardelli, M., Kim, K.W.: 
First-principles analysis of electron-phonon interactions in graphene.
Phys. Rev. B 81,  121412(R) (2010)

\bibitem{CastroNeto09}
Castro Neto, A.H., Guinea, F., Peres, N.M.R., Novoselov, K.S., Geim, A.K.:
The electronic properties of graphene.
Rev. Mod. Phys. 81, 109--162  (2009)

\bibitem{CheianovEtAl2007}
Cheianov, V.V., Fal'ko, V., Altshuler, B.L.:
The focusing of electron flow and a Veselago lens in graphene.
Science 315, 1252--1255  (2007)

\bibitem{Degond_review}
Degond, P.:
Macroscopic limits of the Boltzmann equation: a review.
In: Degond, P., Pareschi, L., Russo, G. (eds.)
Modeling and Computational Methods for Kinetic Equations, pp.\ 3--57. 
Birkh\"auser, Basel (2004)

\bibitem{DEA02}
Degond, P., El Ayyadi, A.:
A coupled Schr\"odinger drift-diffusion model for quantum semiconductor device simulations.
J. Comput. Phys. 181,  222--259 (2002)

\bibitem{DS98}
Degond, P., Schmeiser, C.:
Macroscopic models for semiconductor heterostructures.
J. Math. Phys. 39, 4634--4663 (1998)

\bibitem{DM79}
Duderstadt, J.J., Martin, W.R.:
Transport Theory.
Wiley, New York (1979)

\bibitem{GK95}
Golse, F., Klar, A.:
A numerical method for computing asymptotic states and outgoing distributions for kinetic linear half-space problems.
J. Stat. Phys. 80, 1033--1061 (1995)

\bibitem{KatsnelsonEtAl06}
Katsnelson, M.I.,  Novoselov, K.S., Geim, A.K.:
Chiral tunnelling and the Klein paradox in graphene.
Nat. Phys. 2, 620--625 (2006)

\bibitem{Lee15}
Lee G.H., Park G.H., Lee H.J.:
Observation of negative refraction of Dirac fermions in graphene. 
Nat. Phys. 11, 925--929 (2015)

\bibitem{Lejarreta13}
Lejarreta, j.D.,  Fuentevilla, C.H., Diez, E.,  Cerver\'o J.M.: 
An exact transmission coefficient with one and two barriers in graphene.
J. Phys. A 46, 155304 (2013)

\bibitem{MMR17}
Majorana, A., Mascali, G., Romano, V.:
Charge transport and mobility in monolayer graphene.
J. Math. Ind. 7:4 (2017)

\bibitem{Morandi09}
Morandi, O.:
Wigner-function formalism applied to the Zener band transition in a semiconductor.
Phys. Rev. B 80, 024301 (2009)

\bibitem{MB14}
Morandi, O., Barletti, L.:
Particle dynamics in graphene: collimated beam limit. 
J. Comput. Theor. Transport 43, 1--15  (2014)

\bibitem{Poupaud91}
Poupaud, F.:
Diffusion approximation of the linear semiconductor Boltzmann equation: analysis of boundary layers.
Asymptotic Anal. 4, 293--317 (1991)
 
\bibitem{SW58}
Slonczewski J.C., Weiss, P.R.:
Band structure of graphite. 
Phys. Rev. 109, 272--279 (1958)

\bibitem{Young09}
Young A.F., Kim P.:
Quantum interference and Klein tunnelling in graphene heterojunctions. 
Nat. Phys. 5, 222--226 (2009)

\end{thebibliography}
\end{document}